\documentclass[usegraphix,usenatbib,twocolumn]{mn2e}

\usepackage{amsmath}
\usepackage{amssymb}
\usepackage{times}
\usepackage{graphicx}
\usepackage{subfigure}
\usepackage{epsfig}
\usepackage{txfonts}
\usepackage{hyperref}
\usepackage{paralist}
\usepackage{rotating}

\newcommand{\bs}[1]{\bmath{#1}}

\def\mnras{{MNRAS}}
\def\apj{{ApJ}}
\def\apjs{{ApJS}}

\def\aj{{AJ}}

\def\aap{{AA}}

\def\pra{{PRA}}

\def\nat{{Nature}}

\def\degrees{\,\mathrm{degrees}}
\def\kpc{\,{\rm kpc}}
\def\Fe{{F_{\rm e}}}
\def\kms{{\,\mathrm{kms^{-1}}}}

\def\obs{{\boldsymbol{L}}}
\def\intd{{\mathrm{d}}}

\def\vel{{\boldsymbol{v}}}
\def\params{{\mathcal{P}}}
\def\errmatrix{{\boldsymbol{C}}}

\begin{document}

\pagenumbering{arabic}

\title[On the Alignment of the Second Velocity Moment Tensor in Galaxies] {The
  Alignment of the Second Velocity Moment Tensor in Galaxies} \author[N.W. Evans
  et al.]  {N.~W. Evans$^1$, J.~L. Sanders$^1$, A.~A. Williams$^1$,
  J.~An$^2$, D. Lynden-Bell$^1$, and W.Dehnen$^3$
\medskip
   \\$^1$Institute of Astronomy, University of Cambridge, Madingley Road,
         Cambridge, CB3 0HA, UK
    \\$^2$National Astronomical Observatories, Chinese Academy of
    Sciences, A20 Datun Road, Chaoyang District, Beijing 100012,
    PR~China
    \\$^3$Department for Physics \& Astronomy, University of Leicester,
    Leicester, LE1 7RH, United Kingdom
}
\maketitle

\begin{abstract}
We show that, provided the principal axes of the second velocity moment
tensor of a stellar population are generally unequal and are oriented
perpendicular to a set of orthogonal surfaces at each point, then
those surfaces must be confocal quadric surfaces and the potential
must be separable or St\"ackel. This is true under the mild assumption
that the even part of the distribution function (DF) is invariant
under time reversal $v_i \rightarrow -v_i$ of each velocity component.
In particular, if the second velocity moment tensor is everywhere
exactly aligned in spherical polar coordinates, then the potential
must be of separable or St\"ackel form (excepting degenerate cases
where two or more of the semiaxes of ellipsoid are everywhere the
same). The theorem also has restrictive consequences for alignment in
cylindrical polar coordinates, which is used in the popular Jeans
Anisotropic Models (JAM) of Cappellari (2008).

We analyse data on the radial velocities and proper motions of a
sample of $\sim 7400$ stars in the stellar halo of the Milky Way.  We
provide the distributions of the tilt angles or misalignments from
the spherical polar coordinate systems. We show that in this
sample the misalignment is always small (usually within $3^\circ$) for
Galactocentric radii between $\sim 7$ and $\sim 12$ kpc. The velocity
anisotropy is very radially biased ($\beta \approx 0.7$), and almost
invariant across the volume in our study.  Finally, we construct a
triaxial stellar halo in a triaxial NFW dark matter halo using a
made-to-measure method.  Despite the triaxiality of the potential, the
velocity ellipsoid of the stellar halo is nearly spherically aligned
within $\sim6\degrees$ for large regions of space, particularly
outside the scale radius of the stellar halo.  We conclude that the
second velocity moment ellipsoid can be close to spherically aligned
for a much wider class of potentials than the strong constraints that
arise from exact alignment might suggest.
\end{abstract}


\begin{figure*}
  \centering
  \hspace{-1.5cm}
  \includegraphics[width=7.5in]{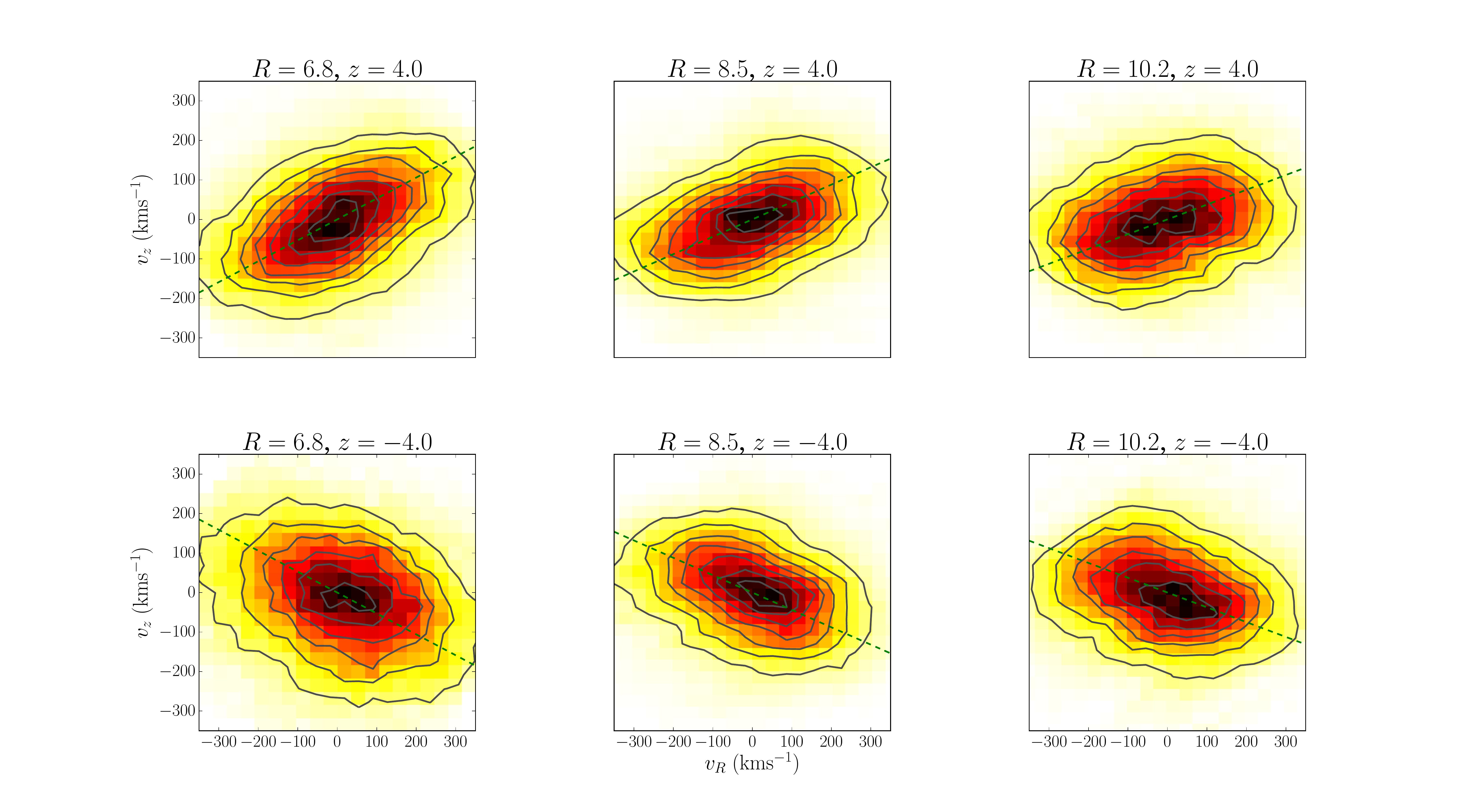}
  \caption{Velocity distributions of our halo sample in $(v_R,v_z)$
    space in 6 spatial bins. The position of each bin centroid $(R,z)$
    is written in kpc above the plot of the velocity distributions in
    that bin.  The green dashed lines are at an angle
    $\mathrm{arctan}\left( z/R \right)$ where $(R,z)$ is the position
    of the bin centroid. The long axis of the $(v_R,v_z)$ distribution
    should coincide with this line if the velocity ellipsoid is
    spherically aligned.}
  \label{fig:SDSSveldists}
\end{figure*}

\begin{figure*}
\centering
  \includegraphics[width=2\columnwidth]{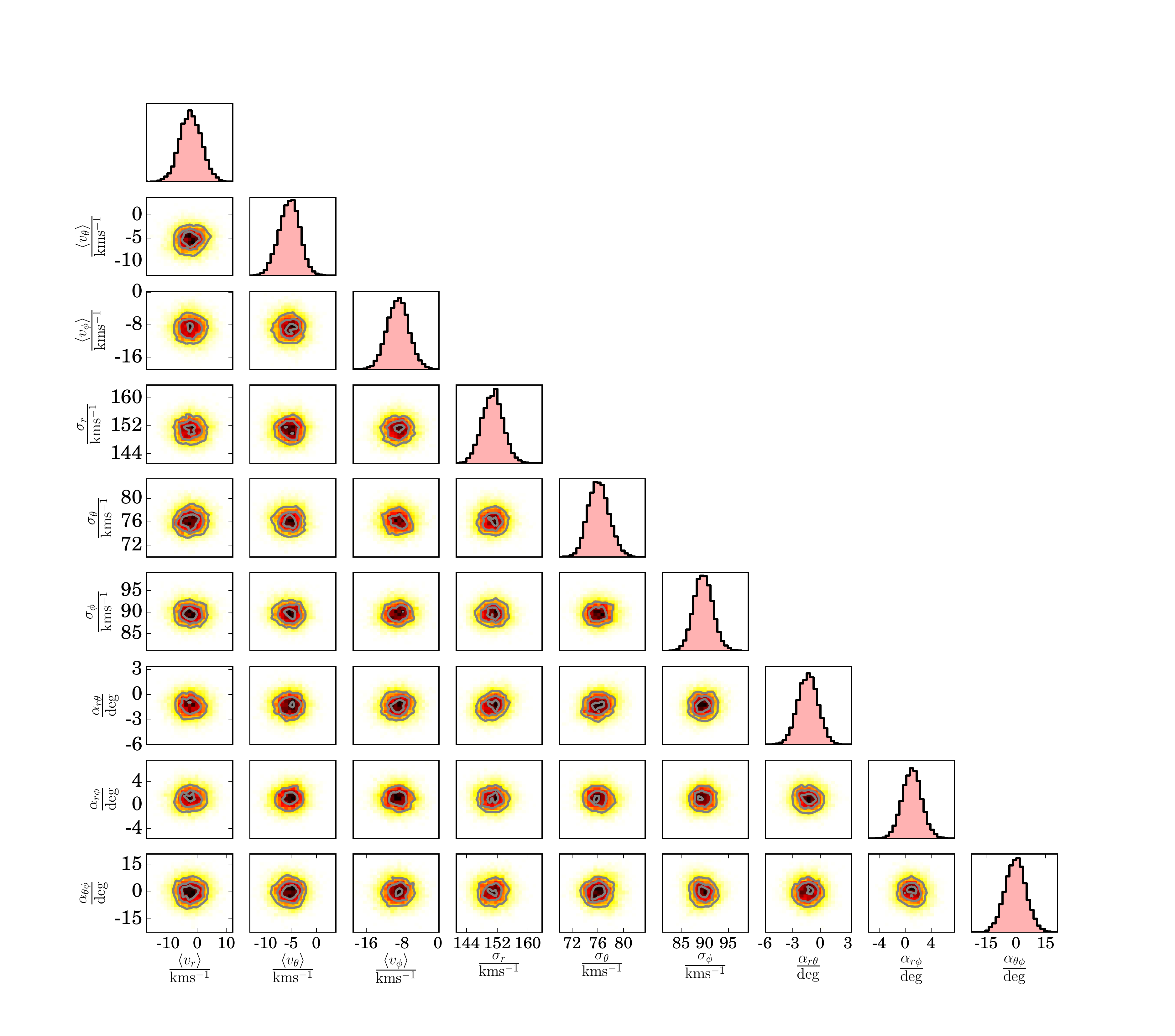}
  \caption{Inference on the parameters of the model described in
    section \ref{sec:model} in the bin $7.7 < R/\kpc < 9.3$, $3 <
    z/\kpc <5$. We can see that there is no noticeable covariance
    between any of the model parameters, and that the tilt angles are
    all consistent with zero at (at least) the $2\sigma$ level.}
  \label{fig:triangleplot}
\end{figure*}

\section{Introduction}

The Jeans equations relate the gravitational potential of a galaxy to
the kinematic poperties of the stars. This is an attractive way to
infer the underlying mass, without the complexities of specifying a
full phase space distribution function (DF).  In this paper, we shall
usually work with the second velocity moment tensor, which is just
\begin{equation}
\langle v_i v_j \rangle = {1\over \rho} \int f v_i v_j d^3 v.
\end{equation}
Here, the subscript indices denote one of the orthogonal coordinate
directions, and the angled brackets represent averaging over the phase
space distribution function $f$, whilst $\rho$ is the density
\citep[see e.g.,][]{Bi08}.  

The second velocity moment tensor is sometimes separated into
contributions from streaming motion and random motion by defining the
velocity dispersion tensor
\begin{equation}
\sigma^2_{ij} \equiv \left \langle (v_i - \langle v_i \rangle) (v_j -
\langle v_j \rangle) \right\rangle.
\end{equation}
In the absence of streaming motions, the second moment tensor and the
velocity dispersion tensor are identical.  The second moment tensor is
a symmetric second-rank tensor and so may always be diagonalized.  The
principal axes of the tensor then form an ellipsoid, which we shall
call the second moment ellipsoid.  In the absence of streaming
motions, this is just the velocity ellipsoid. This paper studies the
alignment of the second moment ellipsoid and its implications for the
underlying gravitational potential.

The Jeans equations are the first-order moments of the collisionless
Boltzmann equation. They are three equations relating the six
independent components of the second velocity moment tensor $\langle
v_i v_j \rangle$ to the density and the potential. Hence, the Jeans
equations cannot uniquely determine the $\langle v_i v_j \rangle$ and
some closure condition must be adopted.  A common choice is the
alignment of the second moment ellipsoid in some coordinate system,
which reduces the number of independent variables from six to three --
namely, the semiaxes of the ellipsoid. For example, \citet{Ca08}
provided an elegant way to solve the Jeans equations, assuming
alignment in the cylindrical polar coordinate system
($R,\phi,z$). These models -- Jeans anisotropic models or JAM -- have
become widely used in analyses of integral field data on elliptical
galaxies~\citep[see e.g.,][]{Ca13}, as well as studies of nuclear
clusters~\citep{Ha11} and lensing galaxies~\citep{vdV10}.
Cylindrically aligned solutions with two of the semiaxes equal (i.e.,
$\langle v_R^2 \rangle = \langle v_z^2 \rangle$) are generated by DFs
depending on the two isolating integrals, energy $E$ and angular
momentum component $L_z$~\citep{Je19}. When all three semiaxes are
different, the validity of the JAM solutions remains unclear. In fact,
\citet{Bi14} has already questioned whether construction of numerical
DFs for models with such properties is possible.

Alignment in the spherical polar coordinate system $(r,\theta,\phi$)
is also often used in the Jeans equations. If two of the semiaxes are
equal (i.e., $\langle v_\theta^2 \rangle = \langle v_\phi^2 \rangle$),
then the spherical aligned solutions can be generated by DFs depending
on the two integrals, energy $E$ and square of the angular momentum
$L^2$. Assuming a spherical density and potential, this ansatz is very
popular as the Jeans equations then reduce to a single equation for
the radial velocity dispersion $\langle v_r^2 \rangle$ together with
an anisotropy parameter $\beta(r)$
\begin{equation}
\beta (r) = 1 -{\langle v_\theta^2 \rangle \over \langle v_r^2
  \rangle}.
\end{equation}
Given a choice for $\beta(r)$, the only non-trivial Jeans equations
can be straightforwardly solved using an integrating factor \citep[see
  e.g.,][]{vdM94, An11, Ag14}. Algorithms for solution of the Jeans
equations using spherical alignment but flattened densities and
potentials have also been developed \citep[see
  e.g.,][]{Ba83,Ba85,EHZ97,Ev15}.  These can in general have three
different principal axes $\langle v_r^2\rangle$, $\langle v_\theta^2
\rangle$ and $\langle v_\phi^2 \rangle$ and so the second moment
ellipsoid is triaxial, but whether they can be realised by physical
(non-negative) DFs remains unclear.

Alignment in spheroidal coordinates ($\lambda,\mu,\phi$) has also been
studied.  This coordinate system is described in, for example,
\cite{MF} or \citet{Bi08}. Within the foci of the coordinate system,
spheroidal alignment approaches cylindrical, whilst at large radii, it
tends to spherical. Spheroidal alignment can therefore be viewed as
interpolating between these two more familiar cases.  It has long been
known that if the gravitational potential is of St\"ackel or separable
form, then the second moment ellipsoid is aligned in spheroidal
coordinates~\citep[see e.g.,][]{Ed15,LB62,ELB}.  However, the
assumption that the ellipsoid is aligned in spheroidal coordinates can
be made without a separable potential~\citep{Ar95}. In fact, this
makes good sense, as models of axisymmetric galaxies using numerical
constructed DFs do suggest the alignment may be close to
spheroidal~\citep[e.g.,][]{De93, Bi14}. Intuitively, for systems like
the stellar halo of the Milky Way, the potential in the inner parts is
controlled by the flattened disk and bulge, in the outer parts by the
rounder dark matter halo, and so again an alignment in spheroidal
coordinates does seems very natural.

Spheroidal coordinates are the axisymmetric limit of ellipsoidal
coordinates $(\lambda,\mu,\nu$). For triaxial St\"ackel models,
\citet{Ed15} already knew that the second velocity moment ellipsoid is
aligned in ellipsoidal coordinates.  The triaxial Jeans equations for
such St\"ackel systems have been studied
sporadically~\citep{LB60,ELB,vdV}. Given the density and potential,
the Jeans equations are now three coupled first-order partial
differential equations for three unknowns, namely $\langle
v_\lambda^2\rangle, \langle v_\nu^2\rangle$ and $\langle v_\mu^2
\rangle$.  Both the prescription of appropriate boundary conditions
and the solution of the equations is challenging. Only very few
general triaxial DFs have ever been numerically constructed. The few
such models available do show approximate alignment in ellipsoidal
coordinates~\citep[e.g.,][]{Sa15}.

Our intention in this paper is to examine what may be legitimately
deduced about the underlying gravitational potential from the
alignment of the second velocity moment ellipsoid.  This is motivated
by arguments of \citet{Sm09a} who claim that: ``If a steady state
stellar population has a triaxial velocity dispersion tensor whose
eigenvectors are everywhere aligned in spherical polar coordinates,
then the underlying gravitational potential must be spherically
symmetric''. This theorem will be examined, slightly corrected, and
then extended below.

This argument was queried by \citet{Bi11}, who created a torus-based
model of the Galaxy that, at locations above the plane of the Milky
Way, possessed a spherically aligned velocity dispersion tensor, even
though the potential was highly flattened. Although \citet{Bi11} did
not provide an explicit counter-example to the theorem of
\citet{Sm09a}, they did question whether in general the potential does
control the tilt.  In their numerically constructed example, they
argued that the tilt of the second velocity moment ellipsoid is
controlled at least as much by the weightings of orbits, and hence the
DF, as compared to the potential. In this respect, \citet{Bi11}
conjectured that insights from separable or St\"ackel models may not
tell the whole story. One of the aims of this paper is to resolve the
tension between these two viewpoints.

The paper is organised as follows. Section 2 studies the one
spheroidal system for which there is hard data on the alignment --
namely, the stellar halo of the Milky Way galaxy.  For elliptical
galaxies, evidence on the alignment is necessarily much more indirect,
as only the line of sight velocity distribution can be measured. We
confirm, and sharpen, the results of \citet{Sm09a} and \citet{Bo10}
that the velocity ellipsoid of stars in the Milky Way stellar halo is
close to spherical or spheroidal alignment.  Section 3 is theoretical
and examines the case of exact spherical alignment. We prove under
quite general conditions that the gravitational potential must be of
separable or St\"ackel form in spherical polars.  This is an
elaboration of the original theorem of \citet{Sm09a}.  The only
exceptions are simpler cases in which the second velocity moment
ellipsoid has either two axes the same and so is a spheroid, or three
axes the same and so is a sphere. Section 4 discusses cylindrical and
spheroidal alignment, and demonstrates, for the first time, that the
assumption of alignment of the (assumed triaxial) second velocity
moment tensor implies separability of the Hamilton-Jacobi equation in
these coordinates.  The converse result -- that the separable or
St\"ackel potentials generate galaxy models in which the second
velocity moment tensor is aligned along the separable coordinate
system -- has been known for some time and is implicit in Eddington's
early work~\citep{Ed15, LB62, Ev11}. Lastly, Section 5 builds models
of stellar haloes in which the second velocity moment tensor is close
to, but not exactly, spherically aligned and tests whether results
remain valid in this approximate regime.  We show that it is possible
to build haloes in flattened potentials in which the velocity
ellipsoid is close to spherical alignment over substantial portions of
configuration space, though not everywhere. However, the more the
potential is flattened, the greater is the magnitude and extent of the
misalignment.

\section{The Stellar Halo of the Milky Way}

\subsection{Background}

The study of the kinematics of stars in the Milky Way halo has been
revolutionized by high quality data derived from the Sloan Digital Sky
Survey (SDSS).  Line of sight velocities are extracted from SDSS
spectroscopy. Proper motions are derived from either multi-epoch SDSS
photometry in Stripe 82~\citep{Br08} or from matches to archival
Schmidt photographic-plate based catalogues~\citep{Mu04}. So, all
three components of velocities are now available for thousands of
halo stars, together with positions and photometric parallaxes.

\citet{Sm09a,Sm09b} constructed a sample of halo subdwarfs using a
reduced proper motion diagram, utilizing the light-motion catalog for
Stripe 82 \citep{Br08}. They extracted a clean sample of 1,782 halo
stars, lying at Galactocentric cylindrical polar radii between 7 and
10 kpc, and at depths of 4.5 kpc or less below the Galactic
plane. They found the first velocity moments to be consistent with
zero to within the statistical error -- see Table 2 of
\citet{Sm09b}. So, the second velocity moments ($\langle v_r^2\rangle,
\langle v_\theta^2\rangle, \langle v_\phi^2\rangle$) are equivalent to
the velocity dispersions ($\sigma_r^2, \sigma_\theta^2,
\sigma_\phi^2$), which we use henceforth in this Section.  They also
found that the velocity ellipsoid of the halo stars is aligned with
the spherical polar coordinate system, with the long axis pointing
towards the Galactic Centre. The halo stars are strongly radially
anisotropic. The semiaxes of the velocity ellipsoid are
($\sigma_r,\sigma_\theta, \sigma_\phi) = (143\pm 2, 77\pm 2, 82\pm 2$)
kms$^{-1}$, which corresponds to an anisotropy parameter
\begin{equation}
\beta = 1 - {\sigma_\theta^2 + \sigma_\phi^2 \over 2 \sigma_r^2}
\end{equation}
of 0.69. They also noted a tentative asymmetry in the $v_\phi$
distribution, but found the $v_r$ and $v_\theta$ distributions
symmetric.

This conclusion was reinforced by the larger sample of
\citet{Bo10}. Here, halo stars are extracted from the SDSS data by
combined colour and metallicity cuts (specifically $0.2 < g-r < 0.4$
and [Fe/H] $< -1.1$).  Requiring the stars possess SDSS spectra for
radial velocities and POSS astrometry for proper motions yields a
large sample of $\sim 7400$ halo stars, with an estimated
contamination of $\sim 6$ per cent. \citet{Bo10} found that the shape
of the velocity ellipsoid is invariant in spherical coordinates within
the volume probed by SDSS and aligned in spherical polar coordinates
(their Figures 12 and 13). They found no statistically significant
tilt from spherical alignment, with deviations modest and ranging
between $1^\circ$ to $5^\circ$. Note that this sample extends over
Galactocentric cylindrical polar radii $6 \lesssim R \lesssim 11$ kpc
and height above the Galactic plane $3\lesssim |z| \lesssim 5$ kpc, and
so is much more extensive than earlier work. Nonetheless, the semiaxes
of the velocity ellipsoid are ($\sigma_r,\sigma_\theta, \sigma_\phi) =
(141 \pm 5, 75\pm 5, 85\pm 5$) kms$^{-1}$, in very good agreement with
\citet{Sm09b}. Bond et al's error bars also include systematic effects
such as errors in the photometric parallaxes, whereas Smith et al's do
not.

Very recently, \citet{Ki15} have examined a still larger sample of
halo stars with Galactocentric radii between 6 and 30 kpc. This
contains blue horizontal branch (BHB) and halo F stars extracted from
SDSS, as well as a sample of still more distant F stars obtained from
Hectospec on the MMT. Although only line of sight velocities are
available, this still can yield constraints on the alignment, as there
are different contributions from both radial and tangential velocities
in different directions on the sky. \citet{Ki15} find the alignment of
the velocity ellipsoid of their halo sample to be close to spherical,
albeit to within quite large uncertainties.

\subsection{Alignment and Symmetries}

Here, we will re-analyse the data of \citet{Bo10}.  We first discuss
how the errors on the observable quantities propagate through to the
alignment.  The errors on the radial velocities are computed by the
SDSS pipeline, to which should be added a systematic error of between
2 kms$^{-1}$ (optimistic) and 6 kms$^{-1}$ (pessimistic). The
proper-motion errors are discussed in \citet{Bo10}, who suggest using
a fixed value of 0.6 mas yr$^{-1}$. We assume that there is zero
covariance between the two proper-motion components.

The SDSS pipeline provides photometric errors, which can be used to
compute random errors in absolute magnitude (assuming that the
inter-band covariances are negligible).  Polynomial relations for the
absolute magnitude of F and G stars as a function of $g$, $r$, $i$ and
metallicity [Fe/H] are given in \citet{Iv08}. We then use Monte Carlo
methods to gain an estimate of the random error in the absolute
magnitude, given the errors in $g$,$r$,$i$ and [Fe/H], and add
0.1 mag in quadrature to account for systematics.  Given the estimates
of error in the absolute magnitude (with systematics included), we
use Monte Carlo methods to estimate the error on the distances of
each star in the sample.

To extract the halo sample, we apply the cuts $0.2 < g - r < 0.6$ and
$-3.0 <$ [Fe/H] $< -1.1$ to the data. We then restrict ourselves to stars 
that are found in the volume $6 < R/\kpc < 11$ and $3 < |z|/\kpc<5$. This 
gives us a final sample of 7418 stars (5006 above the plane and 2412 below).

Figure \ref{fig:SDSSveldists} depicts the velocity distributions of our samples. 
Just through visual inspection, it is clear that the velocity ellipsoid cannot be
aligned with cylindrical coordinates, as the $(v_R,v_z)$ distribution
has a noticeable tilt. The green dashed lines in the plot are at an
angle $\mathrm{arctan}\left(z/R\right)$, the direction that the long
axis of the distribution should point if spherical alignment is
satisfied. By eye, the tilt of the velocity distributions seems to be
in good agreement with spherical alignment. This motivates a
quantitative analysis of the data.

\subsection{Inference on properties of the velocity distributions}
\label{sec:model}

Given three orthogonal velocity components, $(v_1,v_2,v_3)$, the tilt
angle, $\alpha _{ij}$, is the angle between the $i$-axis and the major
axis of the ellipse formed by projecting the three dimensional
velocity ellipsoid onto the $ij$-plane (see e.g., \citet{BM} or
Appendix A of \citet{Sm09a}).
\begin{equation}
\tan(2\alpha_{ij}) =
\frac{ 2\sigma_{ij}^2 }{ \sigma_{ii}^2 - \sigma_{jj}^2 }.
\label{eq:tiltdef}
\end{equation}
Our aim is to diagnose the degree of misalignment in the data. We do
so by using a probabilistic method and a simple model of the velocity
distributions in some coordinate system. We split the data into six
annular bins in $(R,z,\phi)$. Each bin spans the whole range of
$\phi$. There are three bins above the plane and three below it -- in
all cases the vertical range is $3 < |z|/\kpc < 5$.  We then split the
data in to three equally spaced radial bins within the range
$6<R/\kpc<11$. Within each bin, we make the simplifying assumption that
the density distribution in uniform, so that
\begin{equation}
p(R,z,\phi) = \mathcal{U}(R,z,\phi).
\end{equation}
We then model the velocity distribution in each bin as a 3--dimensional 
normal distribution in the three relevant (orthogonal) 
velocity components $(v_1,v_2,v_3)$. Here we focus on the spherical and 
prolate spheroidal cases. The normal distribution 
has mean $\boldsymbol{\mu} = \langle \vel \rangle$ and covariance matrix
\begin{equation}
\boldsymbol{\Sigma} = \begin{bmatrix}
                        \sigma_{1}^2 & \frac{1}{2}(\sigma_{1}^2-\sigma_{2}^2)\tan2\alpha_{12} & \frac{1}{2}(\sigma_{1}^2-\sigma_{3}^2)\tan2\alpha_{13} \\
                        . & \sigma_{2}^2 & \frac{1}{2}(\sigma_{2}^2-\sigma_{3}^2)\tan2\alpha_{23} \\
                        . & . & \sigma_{3}^2
                        \end{bmatrix},
\end{equation}
where $\sigma_i$ are the velocity dispersions and $\alpha_{ij}$ are the
misalignment angles as previously defined, and the matrix is by
construction symmetric. The parameters of the model (per bin) are then
$\params = (\langle \boldsymbol{v} \rangle,
\boldsymbol{\sigma},\boldsymbol{\alpha})$.  We then compute the
likelihood for individual stars within a given bin as
\begin{equation}
p(\obs | \params) = \int \, \intd \obs'\, p(\obs | \obs', \errmatrix)\,p(\obs' | \params),
\end{equation}
where $\obs = (s,l,b,v_{\mathrm{LOS}},\boldsymbol{\mu})$. We assume
Gaussian errors in the data, with some covariance matrix $\errmatrix$
that is diagonal -- which is tantamount to assuming that errors on the
observed quantities are not correlated. Thus, $p(\obs | \obs',
\errmatrix)$ is a 6--dimensional normal distribution with a diagonal
covariance matrix, although in practise we assume that the
measurements of $l$ and $b$ are error--free. Our model is expressed in
the coordinate system given by $\boldsymbol{w} =
(R,z,\phi,v_i,v_j,v_k)$
\begin{equation}
p(\boldsymbol{w} | \params) = \mathcal{U}(R,z,\phi)\mathcal{N}(\boldsymbol{v},\langle \boldsymbol{v} \rangle,\boldsymbol{\Sigma}),
\end{equation}
which is related to the distribution $p(\obs' | \params)$ by the Jacobian factor
\begin{eqnarray}
p(\obs' | \params) &=&  p(\boldsymbol{w} | \params) \left| \dfrac{\partial \boldsymbol{w}}{\partial \boldsymbol{\obs'}} \right|, \nonumber \\
           &=&  p(\boldsymbol{w} | \params) \dfrac{s'^{4}\cos b'}{R(s',l',b')},
\end{eqnarray}
where $R(s',l',b')$ is the cylindrical $R$ coordinate expressed in terms of the line--of--sight 
distance and galactic longitude and latitude. Thus the final expression for the likelihood of 
an individual star in a given bin is
\begin{equation}
p(\obs | \params) = \int \, \intd \obs'\, p(\obs | \obs', \errmatrix)\,\mathcal{U}(R',z',\phi')\,\mathcal{N}(\boldsymbol{v}',\boldsymbol{\mu},\boldsymbol{\sigma})\dfrac{s'^{4}\cos b'}{R(s',l',b')}.
\label{eq:likelihood}
\end{equation}
In order to approximate this likelihood, we draw $N$ samples from the
error ellipsoids of each star (i.e. we sample the distribution $p(\obs
| \obs', \errmatrix$)) and evaluate the likelihood via the
Monte--Carlo sum
\begin{equation}
p(\obs | \params) \simeq \frac{1}{N} \sum_{i=0}^{N}
\mathcal{U}({R'}_i,{z'}_i,\phi'_i)\,\mathcal{N}(\boldsymbol{{v}'}_i,\boldsymbol{\mu},\boldsymbol{\sigma})\dfrac{{s'}_i^{4}\cos
  {b'}_i}{R({s'}_i,{l'}_i,{b'}_i)}.
\label{eq:approxlikelihood}
\end{equation}
This is done for $N=100$ per star and the total likelihood is the
product of equation (\ref{eq:approxlikelihood}) for every star in the
bin. The log--likelihood is explored using the {\sc emcee} ensemble
sampler \citep{Fo13}, implemented in {\sc python}. In the end, we
infer 9 parameters for each of our 6 spatial bins (3 mean velocities,
3 velocity dispersions and 3 misalignment angles). Our priors are
uninformative, so that $-45^\circ < \alpha_{ij} < 45^\circ$ and $\sigma_i >
0$.

\subsection{Analysis in spherical coordinates}

\begin{table*}
  \begin{tabular}{lrrrrrrrrrrrrrrrrrc}
\hline
\hline
\multicolumn{1}{c}{($R,z$)} & \multicolumn{1}{c}{$\langle v_r \rangle$} &
\multicolumn{1}{c}{$\langle v_\theta \rangle$} &
\multicolumn{1}{c}{$\langle v_\phi \rangle$} &
\multicolumn{1}{c}{$\sigma_r$} &
\multicolumn{1}{c}{$\sigma_\theta$} &
\multicolumn{1}{c}{$\sigma_\phi$} &
\multicolumn{1}{c}{$\alpha_{r\theta}$} &
\multicolumn{1}{c}{$\alpha_{r\phi}$} & 
\multicolumn{1}{c}{$\alpha_{\theta\phi}$} \\
\multicolumn{1}{c}{(kpc)} &
\multicolumn{1}{c}{($\kms$)} &
\multicolumn{1}{c}{($\kms$)} &
\multicolumn{1}{c}{($\kms$)} &
\multicolumn{1}{c}{($\kms$)} &
\multicolumn{1}{c}{($\kms$)} &
\multicolumn{1}{c}{($\kms$)} &
\multicolumn{1}{c}{(deg)} & 
\multicolumn{1}{c}{(deg)} &
\multicolumn{1}{c}{(deg)} &
\\
\hline
      (6.85,4.00)   & $-4.3\pm4.2$ & $-4.7\pm2.3$ & $-8.6\pm2.9$ & $161.2\pm3.1$ & $88.7\pm1.8$ & $106.8\pm2.3$ & $-0.06\pm1.3$ & $4.9\pm2.0$ & $5.6\pm4.3$ \\
      (8.50,4.00)   & $-2.2\pm3.7$ & $-5.4\pm2.0$ & $-8.9\pm2.4$ & $150.8\pm2.7$ & $76.1\pm1.6$ & $89.7\pm2.0$ & $-1.3\pm1.1$ & $1.0\pm1.5$ & $-0.2\pm5.3$ \\
      (10.15,4.00)  & $-10.2\pm4.5$ & $-2.6\pm2.9$ & $-0.1\pm2.7$ & $161.6\pm3.4$ & $94.2\pm2.1$ & $80.3\pm2.3$ & $11.4\pm1.3$ & $1.3\pm1.3$ & $14.5\pm5.2$ \\
\hline
      (6.85,-4.00)  & $10.5\pm8.8$ & $5.5\pm5.9$ & $3.3\pm6.5$ & $149.3\pm7.1$ & $73.3\pm4.8$ & $103.2\pm5.2$ & $2.5\pm3.0$ & $-12.7\pm4.9$ & $7.8\pm6.4$ \\
      (8.50,-4.00)  & $-9.7\pm5.4$ & $-4.6\pm3.1$ & $-7.0\pm3.6$ & $140.3\pm4.3$ & $74.6\pm2.6$ & $86.4\pm2.8$ & $2.6\pm1.8$ & $-0.3\pm2.3$ & $3.4\pm8.0$  \\
      (10.15,-4.00) & $-5.0\pm4.5$ & $-2.6\pm2.6$ & $0.4\pm3.2$ & $143.1\pm3.4$ & $69.0\pm2.2$ & $76.4\pm3.0$ & $5.5\pm1.5$ & $2.0\pm1.8$ & $-4.8\pm12.8$ \\
\hline
  \end{tabular}
  \caption{At each location ($R,z$), the inferred values and $1\sigma$ confidence intervals for each of the model parameters.}
  \label{table:parameter_inferences}
\end{table*}

Table \ref{table:parameter_inferences} gives the inferred values of
the model parameters in each of our bins, as well as the $1\sigma$
confidence intervals. In general, we detect no strong correlations
between any of the parameters -- as an example, Figure
\ref{fig:triangleplot} shows the one and two--dimensional projections
of the posterior probability distribution in one of our bins.

\begin{figure*}
\centering
  \includegraphics[width=2\columnwidth]{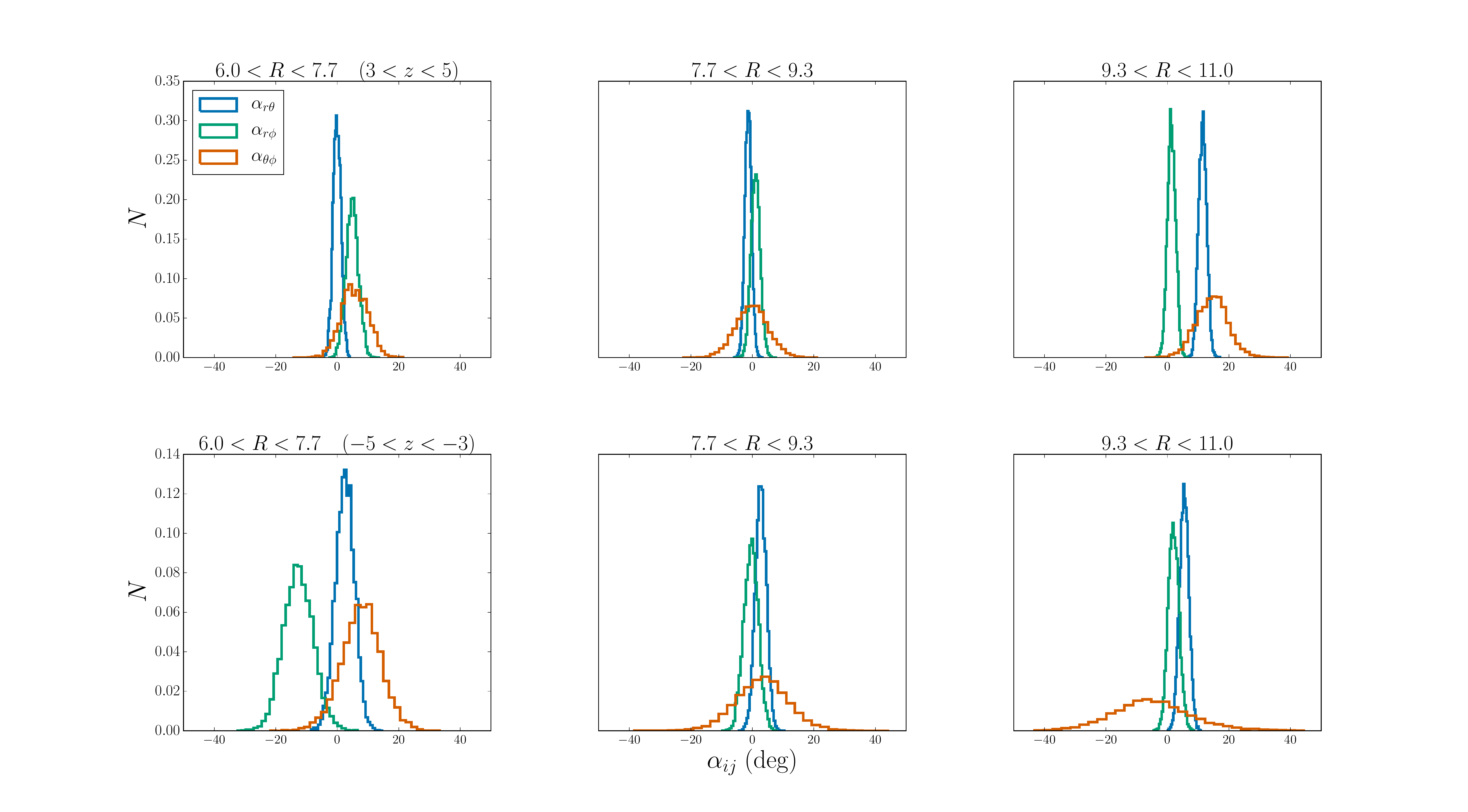}
  \caption{The marginalised posterior distributions on the misalignment 
  angles in each bin. In most of the bins, the three angles are consistent 
  with zero at the $2\sigma$ level or better. The biggest discrepancy is 
  in $\alpha_{r\theta}$ in the above--plane bin spanning $9.3<R/\kpc<11.0$ (top 
  right panel). The angle with the largest uncertainty is always $\alpha_{\theta\phi}$, 
  since $\sigma_\theta$ and $\sigma_\phi$ are the most similar of the three 
  velocity dispersions.}  
  \label{fig:alphas}
\end{figure*}

\begin{figure*}
\centering
  \includegraphics[width=2\columnwidth]{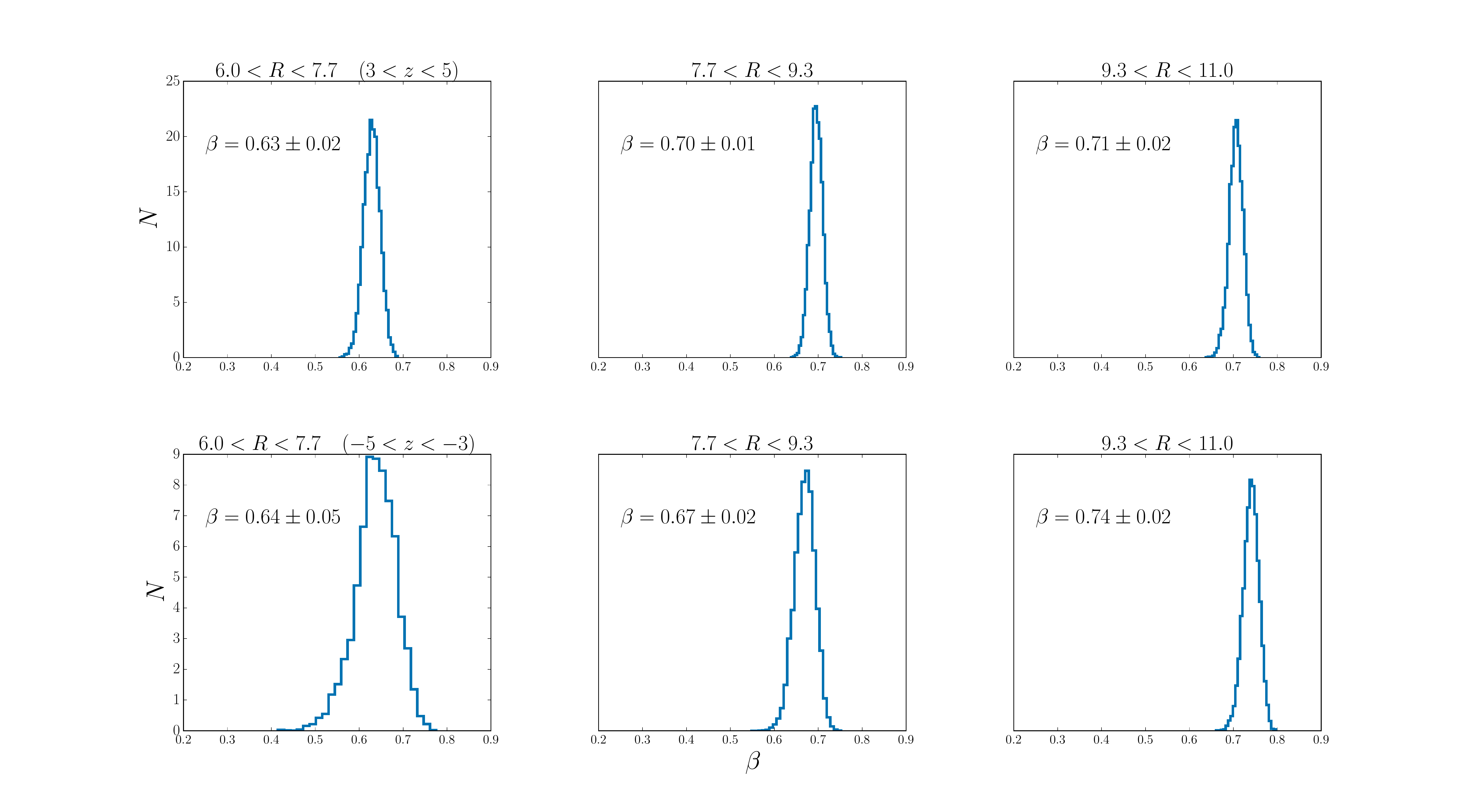}
  \caption{The distributions of the spherical anisotropy parameter
    calculated using the samples from our MCMC chains. The velocity
    anisotropy is very radially biased, and almost invariant across
    the volume in our study.}
  \label{fig:beta}
\end{figure*}

We find that the misalignment angles are small, and that the velocity
dispersions obey $\sigma_r > \sigma_\phi > \sigma_\theta$ --
consistent with a radially biased, flattened stellar halo. Figure
\ref{fig:alphas} gives the one--dimensional marginalised distributions
on the misalignment angles in each of our bins, and Figure
\ref{fig:beta} is the same but for the anisotropy parameter. The
overall picture is of an extremely radially biased population, with
$\beta \sim 0.7$ almost invariant over the whole volume that we probe,
and velocity distributions that are close to spherically aligned --
consistent with the findings of \citet{Bo10} and
\citet{Sm09a,Sm09b}. The inferred misalignment angles are usually
$\sim 3^\circ$ with uncertainties of a similar size, although there
are a few cases where the misalignment is much larger. The most
discrepant case is the bin with edges $9.3 < R/\kpc < 11.0$ and $3 <
z/\kpc <5$, where $\alpha_{r\theta} = 11.4\pm1.3 ^\circ$. Curiously,
this bin is also the only of the six where $\sigma_\theta >
\sigma_\phi$, which is perhaps suggestive that the contamination from
either substructure or the thick disc is more significant in this bin.

We infer non--zero first--order moments in each of the velocity
components. In the following sections, we shall see that the
assumption of symmetry about $\vel = 0$ is necessary to relate the
orientation of the velocity ellipsoids to properties of the matter
distribution. Although it is the case that the first moments are often
discrepant with zero at the one (or sometimes two) sigma level through
inspection of the Monte--Carlo samples of posterior distributions, we
do not believe that the results are truly significant.  The
first--order moments will be the most severely affected of our
parameters by incorrect assumptions about the circular velocity at the
solar position, as well as the velocity of the sun with respect to the
LSR. In all of our analysis, we used the same values for these
velocities as in \citet{Bo10}, and so uncertainties on them are not
reflected in our posterior distributions. Given the fact that the
first moments we infer are always $\sim 10 \kms$ or smaller, it is
highly plausible that incorporating uncertainty on the circular speed
and solar motion could account for these discrepancies from zero.

\subsection{Spheroidal coordinates}

Another coordinate system of interest is the prolate spheroidal system
$(\lambda,\mu,\phi)$, often used in modelling oblate mass
models~\citep[see e.g.,][]{DZ85,De93}.  These coordinates (which are
discussed in greater detail in Section \ref{sec:spheroidalign}) are
related to cylindrical polars via the set of equations
\begin{eqnarray}
\label{eq:spherdef}
R &=& a\,\sinh\lambda\,\sin\mu \nonumber, \\
z &=& a\,\cosh\lambda\,\cos\mu, \nonumber\\
\phi &=& \phi. \\ 
\end{eqnarray}
Surfaces of constant $\mu$ are hyperbolic sheets, whereas surfaces
of constant $\lambda$ are prolate ellipsoids. The quantity $a$ is called
the \textit{focal distance}.  At large distances from the origin,
these surfaces begin to coincide with spherical polar coordinates
($\lambda \sim r$ and $\mu \sim \theta$), whereas at small distances
the coordinate surfaces align with cylindrical polars ($\lambda \sim
R$ and $\mu \sim z$). In order to compute properties of the velocity
distributions, we must also express the velocities in this coordinate
system. These are related to the cylindrical velocities via
\begin{eqnarray}
v_\lambda &=& \dfrac{v_R \cosh \lambda \sin \mu + v_z \sinh \lambda \cos \mu}{\sinh ^2 \lambda + \sin ^2 \mu} \nonumber, \\ \nonumber \\
v_\mu &=& \dfrac{v_R \sinh \lambda \cos \mu - v_z \cosh \lambda \sin \mu}{\sinh ^2 \lambda + \sin ^2 \mu}, \\ \nonumber \\
v_\phi &=& v_\phi. \nonumber
\end{eqnarray}
Since there is a free parameter associated with this coordinate
system, $a$, there are many possible different orientations of the
velocity ellipsoids. In order to decide upon a value of $a$ that 
gives the smallest misalignment angles in the distribution of 
$(v_\lambda,v_\mu,v_\phi)$, we employ a method identical to that 
described above, save for three differences. We do not bin the data 
in this case, but instead simply model the velocity distribution as 
invariant across the entire volume. We assume that this distribution 
has zero mean ($\boldsymbol{\mu}=\boldsymbol{0}$). We also alter the 
covariance matrix $\boldsymbol{\Sigma}$ so that it now reads
\begin{equation}
\boldsymbol{\Sigma} = \begin{bmatrix}
                        \sigma_{\lambda}^2 & 0 & 0 \\
                        . & \sigma_{\mu}^2 & 0 \\
                        . & . & \sigma_{\phi}^2
                        \end{bmatrix}.
\end{equation}
This is done because we are aiming for a value of $a$ that minimizes
misalignment, so our model is of a velocity distribution that has
$\boldsymbol{\alpha}=\boldsymbol{0}$ -- this assumption provides the
constraint on $a$. In this case, then, the model parameters are
$\params = (a,\sigma_\lambda,\sigma_\mu,\sigma_\phi)$. If we were to
bin the data, we would infer four parameters per bin, leading to a
total of 24 parameters. By not binning the data, we reduce our
parameter space to 4 dimensions, which takes significantly less
computational time to sample. Despite our rather crude assumption that
the velocity distribution is invariant in the volume we are studying,
we expect this analysis to give a decent estimate of the focal
distance that reduces misalignment.  Again, we use {\sc emcee} to do
the sampling, with uninformative priors that simply ensure that all
four of the model parameters are greater than zero.

\begin{figure}
\centering
  \includegraphics[width=\columnwidth]{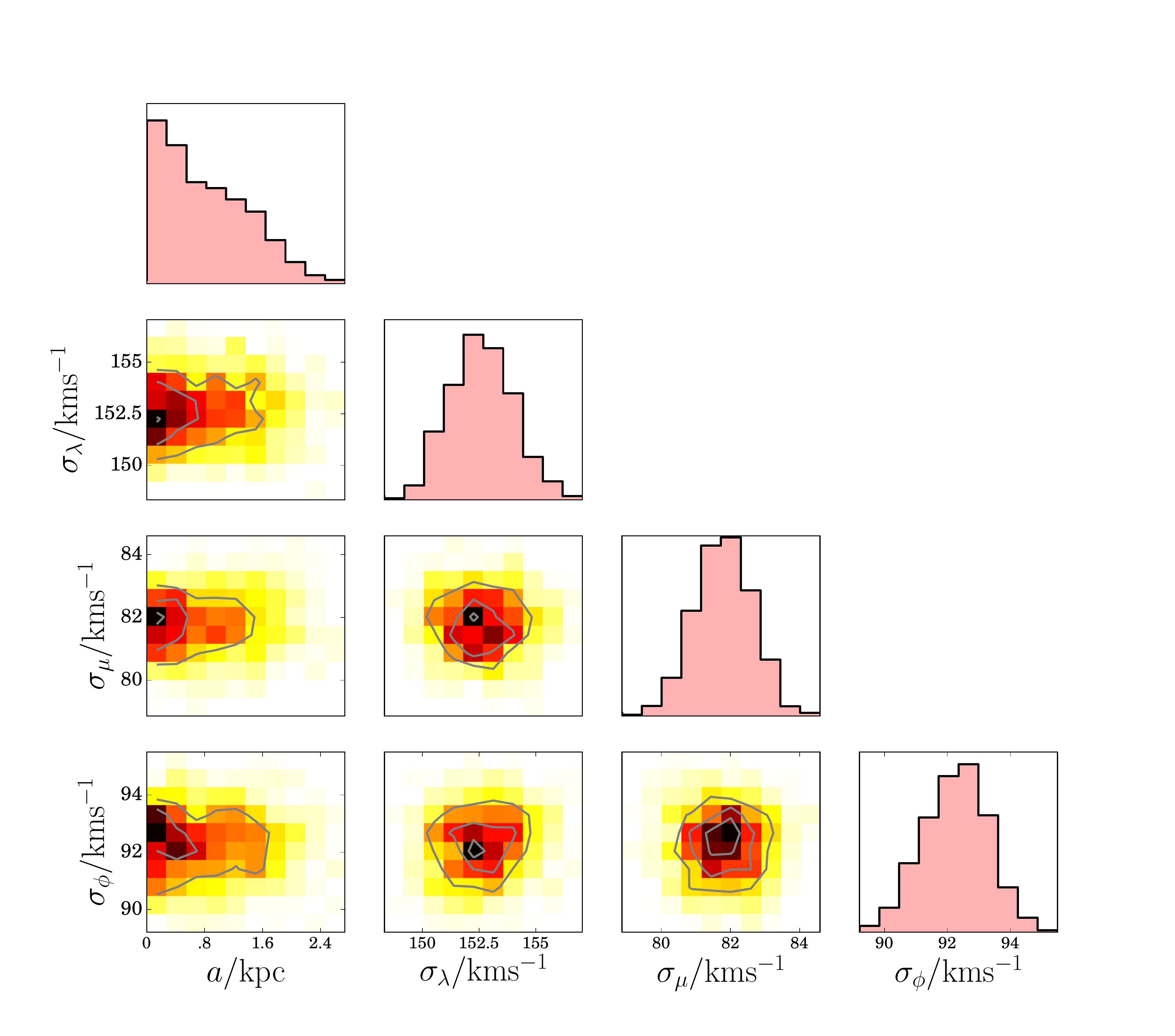}
  \caption{Inference on the best--fit spheroidal focal distance $a$ and 
  the velocity dispersions of the data. One can see that small foci are 
  favoured, so that the spheroidal coordinate system essentially coincides 
  with spherical polar coordinates.}  
  \label{fig:triangleplot_delta}
\end{figure}

Figure \ref{fig:triangleplot_delta} depicts the posterior probability
distributions on the four parameters of the model. In particular, it
is clear that the data favours coordinate systems that have a small
focal distance -- we infer $a=0^{+0.8}_{-0}\kpc$. In this limit, the prolate spheroidal coordinates
coincide with spherical polars. The data, then, suggest that the extra
degree of freedom provided by the prolate spheroidal coordinates does
not produce better alignment of the velocity ellipsoid.

\section{Spherical Alignment}

Motivated by our analysis of the stellar halo, let us assume that the
second velocity moment ellipsoid is aligned with the spherical polar
coordinate system. This means that all the cross terms $\langle v_r
v_\theta \rangle =\langle v_r v_\phi \rangle = \langle v_\theta v_\phi
\rangle$ vanish.  The DF depends on phase space coordinates and so may
have the completely general form $F({\bmath v},{\bmath x}) = F(v_r,
v_\theta, v_\phi, r, \theta, \phi)$. However, the DF is an integral of
motion by Jeans Theorem, and so the Poisson bracket $\{F({\bmath v},
{\bmath x}),H\}$ must vanish. Under time reversal, integrals of motion
remain integrals of motion, so the Poisson bracket $\{F(-{\bmath v},
{\bmath x}),H\}$ must also vanish. Therefore, we can always construct
the even function $\Fe = \frac{1}{2} [F({\bmath v},{\bmath x}) +
  F(-{\bmath v},{\bmath x})]$, which must depend on even powers of the
velocities.  This is of course just the even part of the DF.

The even part of the DF may contain terms like $v_r^l v_\theta^m
v_\phi^n$ provided $l+m+n$ is itself even. However, the cross-terms
$\langle v_r v_\theta \rangle =\langle v_r v_\phi \rangle = \langle
v_\theta v_\phi \rangle$ must all vanish at every position.  On the
grounds of naturalness, we expect that the DF must be of form $\Fe =
\Fe(v_r^2, v_\theta^2, v_\phi^2, r, \theta, \phi)$ and so it must
depend only on even powers of the velocity components in the aligned
coordinate system. This is a key step in the proof, though the
condition can be recast in terms of the fourth velocity moments,
albeit with some effort~\citep{An15}. However, here we can appeal
directly to the data -- we have shown in the preceding section that
the mean motions of velocity distributions of the stellar halo are
consistent with zero in almost every bin. The velocity distributions
are symmetric to a good approximation under the transformations $v_r
\rightarrow -v_r$ or $v_\theta \rightarrow -v_\theta$ or $v_\phi
\rightarrow -v_\phi$ (see e.g., Fig.~\ref{fig:SDSSveldists}). In other
words, without changing the underlying potential, we can always
construct the even part of the DF, which must depend on even powers
only of the individual velocity components.  \citet{Sm09a} reached the
same conclusion that the DF may be chosen to depend only on even
powers, but their line of reasoning was incorrect.

Introducing canonically conjugate momenta $p_r = v_r$, $p_\theta = r
v_\theta$ and $p_\phi = r\sin \theta v_\phi$~\citep{LL}, then the
Hamiltonian is
\begin{equation}
H = {1\over 2} \Bigl( p_r^2 + {p_\theta^2 \over r^2} + {p_\phi^2 \over r^2
  \sin^2 \theta} \Bigr) - \psi(r,\theta,\phi),
\end{equation}
where $\psi$ is the gravitational potential. Without loss of
generality, the DF can now be recast as $\Fe(H, p_\theta^2, p_\phi^2,
r, \theta, \phi)$ by using the Hamiltonian to eliminate $p_r^2$. The
Poisson bracket $\{\Fe,H\}$ must vanish. The $H$ in the DF may be
treated as a constant when evaluating the derivatives in the Poisson
bracket, yielding:
\begin{equation}
p_r{\partial \Fe \over \partial r} +{p_\theta\over r^2}{\partial \Fe \over
  \partial \theta} +{p_\phi \over r^2\sin^2\theta}{\partial \Fe \over
  \partial \phi}
+\left( {p_\phi^2 \cos\theta \over r^2 \sin^3 \theta} + {\partial \psi
  \over \partial \theta} \right) {\partial \Fe \over \partial p_\theta}
+ {\partial \psi \over \partial \phi} {\partial \Fe \over \partial
  p_\phi} =0
\label{eq:sphpolP}
\end{equation}
The first term is antisymmetric in $p_r$, whereas all other terms are
symmetric. So, it must be the case that $\Fe$ does not depend on
$r$. (This is analogous to the elimination of the nodes in the N-body
problem which similarly reduces the order of the system in phase
space by two -- see e.g., \citet{Bo96}.)

We now multiply by $r^2$ and differentiate with respect to $r$ at
constant $H, p_\theta, p_\phi, \theta, \phi$ to obtain:
\begin{equation}
{\partial \Fe \over \partial p_\theta} {\partial^2 \over \partial r
  \partial \theta} \left(r^2\psi\right) + {\partial \Fe \over \partial
  p_\phi} {\partial^2 \over \partial r \partial \phi}
\left(r^2\psi\right) = 0
\end{equation}
The first term is odd with respect to $p_\theta$, the second term odd
with respect to $p_\phi$. However, the equation must remain true under
$p_\phi \rightarrow -p_\phi$ or $p_\theta \rightarrow -p_\theta$, and
so each term must separately vanish.

There are four possibilities -- either (case i)
\begin{equation}
{\partial \Fe \over \partial p_\phi} = 0 = {\partial \Fe \over \partial
  p_\theta},
\label{eq:spherfirst}
\end{equation}
or case (ii)
\begin{equation}
{\partial \Fe \over \partial p_\theta} = 0 = {\partial^2  \over \partial
  r\partial \phi } \left (r^2\psi \right),
\label{eq:sphersecond}
\end{equation}
or case (iii)
\begin{equation}
{\partial \Fe \over \partial p_\phi} = 0 = {\partial^2  \over \partial
  r \partial \theta } \left (r^2\psi \right),
\label{eq:spherthird}
\end{equation}
or case (iv)
\begin{equation}
{\partial^2 \over \partial r \partial \phi} \left( r^2\psi \right)
 =0= {\partial^2 \over \partial r \partial \theta} \left( r^2\psi \right),
\label{eq:spherfourth}
\end{equation}
Cases (i)-(iii) lead to the degenerate cases in which two or three of
the semiaxes of the second velocity moment ellipsoid are the same. For
example, case (i) tells us that $\Fe$ is independent of both
$p_\theta$ and $p_\phi$. On returning to eq~(\ref{eq:sphpolP}), we see
that it also follows that $\Fe$ is independent of the conjugate
coordinates $\theta$ and $\phi$ as well, leaving us with $\Fe
=\Fe(H)$, or solutions with completely isotropic velocity
ellipsoids. The other degenerate cases are when the DF is given by
$\Fe= \Fe(H,|\bmath{L}|)$ in a spherical potential, or $\Fe=
\Fe(E,L_z)$ in an axisymmetric potential. Here, $\bmath{L}$ is the
angular momentum, whilst $L_z$ is the component of $\bmath{L}$ that is
parallel to the symmetry axis. Both lead to the second velocity moment
ellipsoid possessing axial symmetry -- in the former case with
$\langle v_\theta^2 \rangle = \langle v_\phi^2 \rangle$, in the latter
case with $\langle v_r^2 \rangle = \langle v_\theta^2 \rangle$.  These
models have been known since at least the Adams Prize essay of James
Jeans (1919). However, it has also long been known that the velocity
ellipsoid of Milky Way halo stars is triaxial with $\langle v_{r}^2
\rangle > \langle v_\phi^2 \rangle > \langle v_\theta^2 \rangle$
\citep[see e.g.,][]{Wo78, Ch00, Ke07, Sm09b, Bo10}.  Therefore, the
degenerate instances of spherical alignment do not seem to apply to
the case of the Milky Way stellar halo anyhow.

Only case (iv) survives. The solution to $\partial^2 (r^2\psi)/
\partial r\partial \theta = 0$ is
\begin{equation}
r^2\psi = A(r,\phi) + B(\theta,\phi),
\end{equation}
where $A(r,\phi)$ and $B(\theta,\phi)$ are arbitrary functions of the
indicated arguments.  If we also demand that $\partial^2 (r^2\psi)/
\partial r\partial \phi = 0$, then we have
\begin{equation}
{\partial^2 A \over \partial r\partial \phi} =0 \implies A = A_1(r) +
A_2(\phi),
\end{equation}
where $A_1(r)$ and $A_2(\phi)$ are again arbitrary. Thus absorbing
$A_2$ into a new function $A_3(\theta,\phi)$, we obtain the result
\begin{equation}
\psi = {A_1(r)\over r^2} + {A_2(\theta,\phi) \over r^2} = \psi_1(r)
+ {A_2(\theta,\phi) \over r^2},
\label{eq:partressp}
\end{equation}
where $\psi_1(r) = A_1(r)/r^2$ and $A_2(\theta,\phi)$ is an arbitrary
function. So, the radial coordinate separates from the other
coordinates.  This means that there is an additional integral of the
motion by separation of the Hamilton-Jacobi equation:
\begin{equation}
I = {1\over 2}\left(p_\theta^2 + {p_\phi^2\over \sin^2 \theta}\right)
-A_2(\theta,\phi).
\end{equation}
So, we can additionally replace $p_\theta^2$ with $I$ in the DF, which
means that $\Fe = \Fe(H, I, p_\phi^2, r,\theta,\phi)$. Taking the
Poisson bracket and exploiting the fact that both $H$ and $I$ may be
taken as constants during the differentiation, we deduce that $\Fe$
must now be independent of both $r$ and $\theta$, and so are just left
with
\begin{equation}
{p_\phi \over r^2\sin^2\theta}{\partial \Fe \over
  \partial \phi}
+ {\partial \psi
  \over \partial \phi} {\partial \Fe \over \partial p_\phi}
=0
\end{equation}
Multiplying through by $\sin^2\theta$ and differentiating with respect
to $\theta$ gives
\begin{equation}
{\partial^2
\over \partial \theta \partial \phi} \left(\sin^2\theta\,\psi \right)
 =0.
\label{eq:pdesp}
\end{equation}
Inserting eq.~(\ref{eq:partressp}) into eq.~(\ref{eq:pdesp}), this
leads to the separable or St\"ackel potential in spherical polars,
namely
\begin{equation}
\psi = \psi_1(r) + {\psi_2(\theta)\over r^2} + {\psi_3(\phi)\over
  r^2\sin^2\theta},
\label{eq:spherb}
\end{equation}
where $\psi_2(\theta)$ and $\psi_3(\phi)$ are arbitrary functions.
This is the result claimed by \citet{Sm09a}; namely, that if the
second velocity moment ellipsoid is aligned everywhere in spherical
polar coordinates, then the only non-singular potential is spherically
symmetric.  It is worth emphasizing the very general nature of the
assumptions required to deduce the result. {\it Nothing about the
  quadratic nature of the integrals or the separability of the
  underlying potential has been assumed. Rather, these are logical
  consequences that follow from the assumption of spherical
  alignment.} This proof follows the outline of the one presented in
\citet{Sm09a}, amplifying the working where necessary. It differs in
one respect. \citet{Sm09a} assumed that $\langle v_r \rangle =0$
implies that the DF depends on $v_r^2$. This though is not necessarily
true.

Note too that for the purposes of the theorem it is immaterial whether
the population self-consistently generates the gravitational field or
not. The result holds good for tracer populations moving in an
externally imposed potential, as well as populations that generate the
gravity field in which they move.  Finally, although the singularities
in the potentials~(\ref{eq:partressp}) or (\ref{eq:spherb}) at $r=0$
may seem objectionable, we show in Appendix A that this awkwardness
can sometimes be avoided.

\section{Other Alignments}

Other alignments of the second velocity moment ellipsoid are also of
interest in galactic astronomy and dynamics. Here, we consider
cylindrical and spheroidal alignment in some detail. From our work on
spherical alignment, we may conjecture that the only possible
solutions for triaxial velocity ellipsoids are the separable or
St\"ackel potentials. We now demonstrate that such is indeed true.

\subsection{Cylindrical Alignment}

This case is interesting because it has has implications for the
popular JAM models introduced by \citet{Ca08}, which assume
cylindrical alignment. In JAM, the only non-vanishing components of
the second velocity moment tensor are $\langle v_R^2 \rangle$,
$\langle v_\phi^2 \rangle$ and $\langle v_z^2 \rangle$.  JAM models
assume a fixed anisotropy $\beta_z = \langle v_z^2\rangle/\langle
v_R^2\rangle$. Using the boundary conditions that the velocity moments
vanish at infinity then leads to an elegant way of solving the Jeans
equations for the velocity dispersion as quadratures.

We take as our starting point the fact that the second velocity moment
ellipsoid is aligned with the cylindrical polar coordinate system so
that all the cross terms $\langle v_R v_\phi \rangle =\langle v_z
v_\phi \rangle = \langle v_R v_z \rangle$ vanish. As before, we take
the even part of the DF to be $\Fe = \Fe(p_R^2, p_\phi^2, p_z^2, r,
\phi, z)$, where the canonically conjugate momenta are $p_R = v_R$,
$p_\phi= R v_\phi$ and $p_z = v_z$. The Hamiltonian is
\begin{equation}
H = {1\over 2} \Bigl( p_R^2 + {p_\phi^2 \over R^2} + p_z^2 \Bigr) -
\psi(R,\phi,z),
\end{equation}
where $\psi$ is the gravitational potential. The even part of the DF
can now be recast as $\Fe(p_R^2, p_\phi^2, H, R, \phi,z)$ by using the
Hamiltonian to eliminate $p_z^2$. The $H$ in the DF may be treated as
a constant when evaluating the derivatives in the Poisson bracket,
yielding:
\begin{equation}
p_R{\partial \Fe \over \partial R} +{p_\phi \over R^2}{\partial \Fe \over
  \partial \phi} +p_z {\partial \Fe \over
  \partial z}
+ \left({p_\phi^2\over R^3} +{\partial \psi \over \partial R }\right) {\partial \Fe \over \partial
  p_R}
+ {\partial \psi
  \over \partial \phi} {\partial \Fe \over \partial p_\phi}
=0
\label{eq:cylpolP}
\end{equation}
As only the third term involves $p_z$, it follows that ${\partial \Fe
  / \partial z} =0$ and so $\Fe$ must also be independent of $z$. Now
differentiate with respect to $z$ to obtain
\begin{equation}
{\partial \Fe \over \partial p_R} {\partial^2 \psi \over \partial R
  \partial z} + {\partial \Fe \over \partial
  p_\phi} {\partial^2 \psi \over \partial \phi \partial z} = 0
\end{equation}
The first term is odd with respect to $p_R$, the second term odd
with respect to $p_\phi$. Hence, for this to be generally true, each term
must separately vanish.

There are again four possibilities -- either (case i)
\begin{equation}
{\partial \Fe \over \partial p_R} = 0 = {\partial \Fe \over \partial
  p_\phi},
\label{eq:cylfirst}
\end{equation}
or case (ii)
\begin{equation}
{\partial \Fe \over \partial p_R} = 0 = {\partial^2 \psi \over \partial \phi \partial z},
\label{eq:cylsecond}
\end{equation}
or case (iii)
\begin{equation}
{\partial \Fe \over \partial p_\phi} = 0 = {\partial^2 \psi  \over \partial
  R\partial z },
\label{eq:cylthird}
\end{equation}
or case (iv)
\begin{equation}
{\partial^2 \psi \over \partial \phi \partial z}
 =0= {\partial^2 \psi \over \partial R \partial z}
\label{eq:cylfourth}
\end{equation}
As before, cases (i) - (iii) provide the degenerate solutions in which
one or more of the second velocity moments is everywhere the same.
This leaves case (iv), which is non-degenerate and $\langle v_R^2
\rangle$, $\langle v_\phi^2 \rangle$ and $\langle v_z^2 \rangle$ are
all unequal, as the general case.  Solving $\partial^2\psi/ \partial
\phi \partial z = 0$ is
\begin{equation}
\psi = A(R,\phi) + B(R,z),
\end{equation}
where $A(R,\phi)$ and $B(R,z)$ are arbitrary functions of the
indicated arguments.  When we also demand that $\partial^2 \psi/
\partial R\partial z = 0$, then we have
\begin{equation}
\psi(R,\phi,z) = A(R,\phi) + \psi_3(z),
\label{eq:partres}
\end{equation}
where $\psi_3(z)$ is an arbitrary function. Notice that we have now
proved that the $z$ component separates from the other coordinates,
and that the energy in the $z$ direction $H_z$
\begin{equation}
H_z = {1\over 2} p_z^2 - \psi_3(z),
\end{equation}
is an integral of motion.

Of course, we can now repeat our calculation by recasting the even
part of the DF as $\Fe(H, p_\phi^2, H_z, R, \phi, z)$ using the
Hamiltonian $H$ to eliminate $p_R^2$ and $H_z$ to eliminate
$p_z^2$. Again taking the Poisson bracket $\{\Fe,H \}=0$, we
straightforwardly establish that $\Fe$ must be independent of the
conjugate coordinates $R$ and $z$, and are left with
\begin{equation}
{p_\phi \over R^2}{\partial \Fe \over
  \partial \phi}
+ {\partial \psi
  \over \partial \phi} {\partial \Fe \over \partial p_\phi}
=0.
\end{equation}
Multiplying through by $R^2$ and differentiating with respect to $R$
gives
\begin{equation}
{\partial^2 (R^2\psi) \over \partial R \partial \phi} =0.
\end{equation}
So, using eq~(\ref{eq:partres}), the final solution for the potential is
\begin{equation}
\psi = \psi_1(R) + {\psi_2(\phi) \over R^2} + \psi_3(z),
\end{equation}
where $\psi_1(R)$ and $\psi_2(\phi)$ are arbitrary.  This is the separable or
St\"ackel potential in cylindrical polars~\citep[see e.g.,][]{LL,Go80}.

Of course, it has long been known that if the potential is separable
in cylindrical polars, then the second velocity moment ellipsoid is
aligned in the cylindrical polar coordinate system. What has been
proved here for the first time is the converse. If the second velocity
moment ellipsoid is non-degenerate and aligned in cylindrical polar
coordinates, then the gravitational potential is separable or
St\"ackel in cylindrical polar coordinates. The degenerate cases are
the ones in which at least two of the semiaxes are everywhere the
same, and so correspond to models with DFs that have $F(E)$ (isotropic
DFs), $F(E,L_z)$ (axisymmetric potential) and $F(E, E_z)$
(translationally invariant potential).

What then is the status of the JAM models of \citet{Ca08}? The Jeans
solutions are cylindrically aligned with (in general) three unequal
axes. If the even part of the velocity distributions is symmetric 
(i.e. invariant under $v_R \rightarrow - v_R$, $v_\phi \rightarrow -
v_\phi$ and $v_z \rightarrow - v_z$), then the only such physical
models must have potentials that separate in the cylindrical polar
coordinate system. Unfortunately, this yields through Poisson's
equation a total matter density of the form
\begin{equation}
\rho(R,\phi,z) = \rho_1(R) + {\rho_2(\phi) \over R^4} + \rho_3(z).
\end{equation}
The fact that the density separates into stratified layers in $z$ with
the same profile in ($R,\phi$) makes this unrealistic for all known
astrophysical objects.  For arbitrary and astrophysically realistic
potentials, the JAM models are unlikely to be physical. The only
loophole is if the even part of the DF does not fulfill the symmetry
requirement. This is shown by \citet{An15} to be equivalent to
requiring the fourth moments $\langle v_Rv_\phi^3\rangle, \langle v_x
v_\phi^3\rangle, \langle v_R^3v_\phi \rangle, \langle v_R^2 v_\phi v_z
\rangle, \langle v_Rv_z^2v_\phi\rangle$ and $\langle v_z^2
v_\phi\rangle $ not to all vanish.

We advocate exercising care in the use of JAM solutions because the
alignment also seems unnatural (except perhaps in the central parts of
elliptical galaxies). Physically, we expect astrophysical objects to
be at least roughly aligned in spherical polar coordinates rather than
cylindrical, as is borne out by our investigations of the stellar
halo.  Probably, it makes sense to use the JAM solutions for
preliminary models only before they are elaborated upon with
Schwarzschild~\citep{Sc79} or Made-To-Measure methods~\citep{Sy96,
  Dehnen2009}.

\subsection{Spheroidal Alignment}
\label{sec:spheroidalign}

It is straightforward to generalize the proof to the instance of
axisymmetric stellar systems with second velocity moments aligned in
spheroidal coordinates ($\lambda,\mu,\phi$). This coordinate system
has been introduced in eq.~(\ref{eq:spherdef}) and is described in
detail in, for example, \cite{MF} or \citet{Bi08}.  Here, we will show
that if the second velocity moment ellipsoid is everywhere aligned in
spheroidal coordinates ($\langle v_\lambda v_\phi \rangle = \langle
v_\lambda v_\mu \rangle = \langle v_\mu v_\phi \rangle = 0$), then the
gravitational potential has St\"ackel or separable form. Again, we
construct the even part of the DF.  Introducing canonical coordinates,
the DF has the form $\Fe(p_\lambda^2, p_\mu^2, p_\phi^2,
\lambda,\mu,\phi)$. The Hamiltonian is
\begin{equation}
H = {1\over 2} \Bigl( {p_\lambda^2\over P^2} + {p_\mu^2 \over Q^2} + {p_\phi^2 \over R^2} \Bigr) - \psi(\lambda,\mu,\phi),
\end{equation}
where the scale factors are $P, Q$ and $R$ are
\begin{eqnarray}
P^2 &=& {\lambda-\mu \over 4(\lambda + a)(\lambda+b)},\qquad
Q^2 = {\mu -\lambda \over 4(\mu + a)(\mu+b)},\nonumber \\
R^2 &=& {(\lambda+a)(\mu+a)\over a-b},
\label{eq:scalefacs}
\end{eqnarray}
and $a$ and $b$ are constants (see for example the Tables of
\citet{LB62} or eq (6) of \citet{ELB} or Section 2 of
\citet{DZ85}). This implies that the DF can be re-written as $\Fe(H,
p_\mu^2, p_\phi^2, \lambda,\mu,\phi)$ Just as before, requiring the
Poisson bracket $\{H,\Fe\}$ to vanish implies that $\partial \Fe /
\partial \lambda$ also vanishes and so $\Fe$ is independent of
$\lambda$. This leaves us with the condition:
\begin{equation}
A {\partial \Fe \over \partial p_\mu} - {\partial\psi \over\partial \phi} {\partial \Fe \over
  \partial p_\phi} = {p_\mu\over Q^2} {\partial \Fe \over \partial
  \mu} + {p_\phi \over R^2} {\partial \Fe \over \partial \phi}
\end{equation}
with
\begin{equation}
A = {1\over \lambda-\mu}\left[
{p_\mu^2\over 2} {\partial \over \partial \mu }\left( {\lambda-\mu \over Q^2} \right)
+ {p_\phi^2\over 2} {\partial \over \partial \mu} \left( {\lambda-\mu \over R^2} \right) - {\partial \over \partial \mu}\left(
(\lambda-\mu) \psi \right) +H\right].
\end{equation}
Again, the equation must hold on transforming $p_\mu \rightarrow
-p_\mu$, so that in the general case (i.e., ignoring degenerate cases
like isotropy), we must have
\begin{equation}
A {\partial \Fe \over \partial p_\mu}
 = {p_\mu\over Q^2} {\partial \Fe \over \partial
  \mu}, \qquad\qquad
{\partial\psi \over\partial \phi} {\partial \Fe \over
  \partial p_\phi}=
{p_\phi \over R^2} {\partial \Fe \over \partial \phi}.
\end{equation}
Multiplying the first equation by $\lambda-\mu$, differentiating
with respect to $\lambda$ at constant $H$ and then using
the definitions of the scale factors gives us the simple result
\begin{equation}
{\partial^2 \over \partial \lambda \partial \mu} \left( (\lambda-\mu)\psi \right) =0.
\end{equation}
Integrating up, this gives us
\begin{equation}
\psi = {A(\lambda,\phi) - B (\mu,\phi) \over \lambda - \mu},
\label{eq:partresspheroid}
\end{equation}
where $A(\lambda,\phi)$ and $B(\mu,\phi)$ are arbitrary.

Now, we can return to the beginning and instead of eliminating
$p_\lambda^2$ in terms of $H$, we can eliminate $p_\phi^2$ so that the
DF is $\Fe(p_\lambda^2, p_\mu^2, H, \lambda,\mu,\phi)$. Repeating the
steps gives us
\begin{equation}
{\partial^2 \over \partial \lambda \partial \phi} \left( R^2\psi
\right) =
{\partial^2 \over \partial \mu \partial \phi} \left( R^2\psi
\right) = 0,
\end{equation}
from which on inserting eq.~(\ref{eq:partresspheroid}), we obtain the
separable or St\"ackel potential in spheroidal coordinates
\begin{equation}
\psi = {f_1(\lambda) - f_2 (\mu) \over \lambda - \mu} + {f_3(\phi),
\over R^2},
\end{equation}
with $f_1(\lambda)$, $f_2(\mu)$ and $f_3(\phi)$ arbitrary functions of
indicated arguments.  Of course, these potentials have a long history
in both classical mechanics \citep[e.g.,][]{Le04,Wh17,We24,Ei34,Ei48}
and stellar dynamics \citep[e.g.,][]{Ed15,Cl37,LB62}.  In astrophysical
applications, it is usual to set $f_3(\phi) = 0$ as otherwise the
gravitational potential diverges on the axis $R=0$. \citet{DZ85}
provided examples -- now known as the perfect oblate or prolate
spheroids -- of realistic self-gravitating stellar systems with
density stratified on similar concentric spheroids that have a
potential of St\"ackel form. The models are important, as their
orbital structure is generic~\citep[see e.g.,][]{Bo96,Bi08}.

It has been known for many years that, in the axisymmetric St\"ackel
potentials, the second velocity moment ellipsoid is aligned in spheroidal
coordinates~\citep[e.g.,][]{LB60, DZ85, ELB}. We have shown here the
converse also holds true. If the velocity ellipsoid is spheroidally
aligned everywhere and the even part of the DF symmetric under
$v_\lambda \rightarrow -v_\lambda, v_\phi \rightarrow -v_\phi$ and $v_\mu
\rightarrow - v_\mu$), then the potential must be of separable or
St\"ackel form.

Spherical polar, cylindrical polar and spheroidal coordinates are all
limits of the most general case, ellipsoidal coordinates~\citep{MF}.
It may well be suspected that the theorem holds true for ellipsoidal
coordinates as well. Such is indeed true, but we relegate details of
this, the most cumbersome case, to Appendix B. Finally, it is also
reasonable to suspect that a general proof can be found, irrespective
of the coordinate system. In Appendix C, we show that just the
assumption of a triaxial velocity ellipsoid aligned everywhere in some
orthogonal coordinate system, together with the existence of a
steady-state DF with symmetries in velocity space, is sufficient to
constrain the system to be St\"ackel in ellipsoidal coordinates, or
one of its limits (see also \citet{An15}). Although this material has
been placed in an Appendix as it is somewhat mathematical, nonetheless
we regard it as a more powerful proof of Eddington's (1915) theorem
which does not rest on the so-called ``ellipsoidal hypothesis'',
namely the assumption that the DF depends on a single quadratic
function of the velocities (see e.g. \citet{Ch42} and \citet{Cu14} for
later work on the ellipsoidal hypothesis).

\section{Made-to-Measure Models of the 
Stellar Halo}

Theorems on exact alignment are interesting, but in practice the
alignment is approximate and the data extend only over high latitude
fields that are comparatively nearby. Is it possible to build models
in which the alignment is close to spherical, but the gravitational
potential is flattened?

To investigate this, we use a made-to-measure method to construct a
triaxial stellar halo tracer population in a potential generated by a
triaxial NFW dark matter halo. We utilise the made-to-measure code of
\citet[][hereafter D09]{Dehnen2009}. The construction of this model
follows very closely the construction of the models presented in D09
and here we only briefly describe the made-to-measure method.

The made-to-measure technique was pioneered by \cite{Sy96}. In their
formulation, an equilibrium model is constructed by evolving an
$N$-body simulation whilst simultaneously adjusting the particle
weights until a merit function is optimized. The merit function is
expressed as
\begin{equation}
Q = \mu S - \frac{1}{2}C,
\end{equation}
where $C$ is a cost function that quantifies the deviation of the
model from our target model (for instance, the $\chi^2$ difference
between the model moments and the target moments) and $S$ is an
entropy term that regularizes the weight distribution of the
particles. The $N$-body simulation is evolved whilst each particle
weight $w_i$ is adjusted according to a first-order differential
equation that maximises the merit function $Q$. One difficulty
encountered by \cite{Sy96} is that the cost function naturally
fluctuates as the simulation is evolved due to Poisson noise, so some
form of smoothing is required to ensure the algorithm converges. D09
pointed out that \citeauthor{Sy96}'s method of simply averaging the
model properties used in the cost function did not ensure the model
converged, so D09 instead proposed smoothing the merit function $Q$ by
making the weight-adjustment equation a second-order differential
equation. Several other improvements to the original algorithm were
presented by D09. In his formulation \begin{inparaenum}\item each
  particle is evolved on its own dynamical timescale such that the
  outer parts of the model converge as rapidly as the inner
  regions, \item a total weight constraint is included as part of the
  merit function and \item the particles are resampled when the ratio
  between the minimum and maximum weight (normalized with respect to
  the priors) exceeds a chosen value\end{inparaenum}. The last of
  these is implemented by, every so often, drawing new particles from
  the original set with probability proportional to their
  prior-normalized weights and adding a random velocity offset with a
  magnitude that declines exponentially as a function of the
  simulation time. For the model presented below we use very similar
  parameters for the algorithm as those presented in D09.

Using the made-to-measure method, we construct an equilibrium tracer
stellar population inside a fixed dark matter potential. The density
profile of both the dark matter and the tracer population is given by
a truncated double power law of the form
\begin{equation}
\rho_i\propto\bigg(\frac{q_i}{r_{0,i}}\bigg)^{-\gamma_i}\bigg[1+\bigg(\frac{q_i}{r_{0,i}}\bigg)\bigg]^{(\gamma_i-\beta_i)}\mathrm{sech}\bigg(\frac{q_i}{r_{t,i}}\bigg),
\end{equation}
where $r_{0,i}$ is the scale radius of the $i$th component, $r_t$ the truncation radius and the elliptical radius $q$ is defined as
\begin{equation}
q_i^2=\frac{x^2}{a_i^2}+\frac{y^2}{b_i^2}+\frac{z^2}{c_i^2},
\end{equation}
with $a_ib_ic_i=1$. We choose the parameters for the NFW halo as
$\gamma_{\mathrm{NFW}}=1$, $\beta_{\mathrm{NFW}}=3$,
$(b/a)_{\mathrm{NFW}}=0.9$, $(c/a)_{\mathrm{NFW}}=0.8$,
$r_{0,\mathrm{NFW}} = r_0$ and $r_{t,\mathrm{NFW}} = 10r_0$, whilst
the target stellar halo has $\gamma_{S}=1$, $\beta_{\mathrm{S}}=4.5$,
$(b/a)_{\mathrm{S}}=0.8$, $(c/a)_{\mathrm{S}}=0.6$,
$r_{0,\mathrm{S}}=1.73r_0$ and
$r_{t,\mathrm{S}}=9r_{0}$. These choices are motivated by
several studies of the halo. In the dynamical models of
\cite{Piffl2014}, the NFW dark matter halo was found to have a scale
radius of $r_0=15.5\kpc$, where the constraint comes mostly from the
mass measurements of \cite{WE1999} and the requirement that the halo
lies on the mass-concentration relation. \cite{De11b} found from a
population of BHB stars that the stellar halo had a scale radius of
$r_{0,\mathrm{S}}=27\kpc$ and a flattening
$(c/a)_{\mathrm{S}}=0.59$. The axis ratios of the NFW profile are
representative of those found in cosmological simulations \citep[see
  for example][]{Bryan2013}.

The cost function corresponding to this density profile is given by
\begin{equation}
C_\rho = \sum_{\bs{n}}\bigg(\frac{A_{\bs{n}}-B_{\bs{n}}}{\sigma_{\bs{n}}}\bigg),
\label{eq:cost}
\end{equation}
where $A_{\bs{n}}$ is the dot product of the model potential with the
$\bs{n}^{th}=(n,l,m)^{\mathrm{th}}$ basis function of the
\cite{Zhao1996} basis set chosen to match the outer slope of the
target model, and $B_{\bs{n}}$ is the corresponding target moment with
$\sigma_{\bs{n}}$ an estimate of the error in the moment calculated
from the variance of $100$ realizations of the target model. We set
$n_\mathrm{max}=20$ and $l_\mathrm{max}=12$ such that $588$ moments
are used to describe the density. Similarly, the potential used to
describe the NFW halo is represented as a basis function expansion
generated from a single sample of the density distribution. $10^6$
particles were used for each realization of the stellar halo and
$10^8$ for realization of the dark matter density distribution. The
total mass of the NFW halo is $M$ and the stellar halo is treated
purely as a tracer population.

Additionally, we impose an anisotropy profile for the tracer
population as
\begin{equation}
\beta(q) = 1-\frac{\sigma_\theta^2+\sigma_\phi^2}{2\sigma_r^2} = \frac{\beta_0+\beta_\infty(q/r_0)}{1+(q/r_0)}
\end{equation}
which goes from $\beta_0$ at small radii to $\beta_\infty$ at large
radii over a scale $r_0$. We choose $\beta_0=0$ and
$\beta_\infty=0.75$ which approximates the form for the anisotropy
found by \cite{WE2015}. The corresponding cost function $C_\beta$ is
the $\chi^2$ deviation of the model anisotropy from the target
calculated in elliptical shells with the same shape as the target
density. The error is estimated as the Poisson noise arising from a
measurement of the anisotropy from samples of an uncorrelated normal
velocity distribution.

The initial $N$-body model is chosen to be the ergodic model with
the required radial density profile, that is then flattened by the appropriate axis ratios and the velocities scaled to satisfy the tensor virial theorem. We began by running this model for $200$ time units subject to the density constraints. The model converged
within $\sim50$ time units and subsequent evolution with the weight
adjustment switched off did not cause the model to deviate
significantly from the target distribution. The anisotropy of this
model was increased from weakly radial ($\beta\sim0.2$)
at $0.1r_0$ to more strongly radial ($\beta\sim0.5$) at $r=r_{t,\mathrm{S}}$. We then
proceeded to evolve the model further subject also to the anisotropy
constraint. The model converged within $\sim30$ time units and again
did not deviate from the target under subsequent evolution with no
weight adjustment.

\begin{figure}
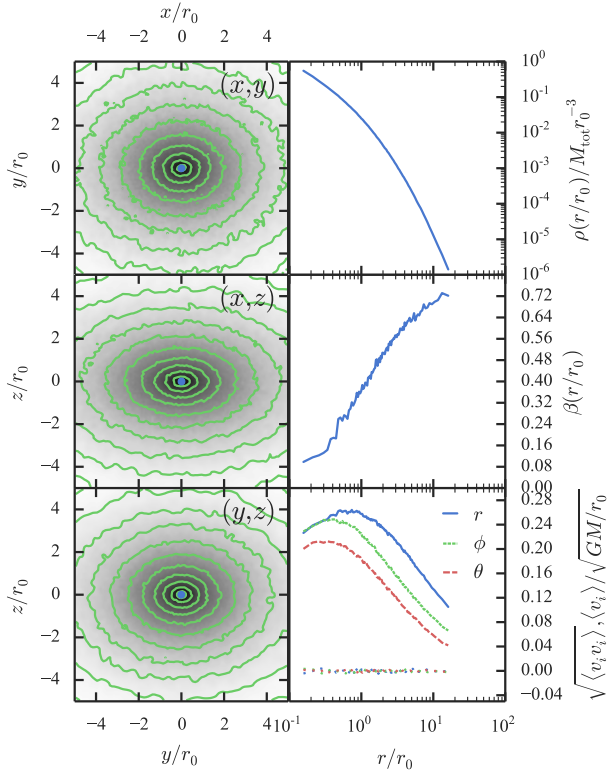

$$\includegraphics[width=\columnwidth]{{{stellarhalo_tracer_beta_nweparams_walterres.nbody}}}$$
\caption{Made-to-measure stellar halo properties: the left panels show
  the density integrated along the line of sight in the three
  principal planes. The green contours are evenly spaced in
  logarithmic density and the blue point is the centre of mass. The
  top right panel shows the spherically-averaged density profile, the
  middle right panel the spherically-averaged anisotropy profile and
  the bottom right panel the spherically-averaged velocity dispersions
  and mean velocities. Note all the mean velocities are zero, as there
  is no net streaming.}
\label{fig::m2m_density}
\end{figure}

\begin{figure}
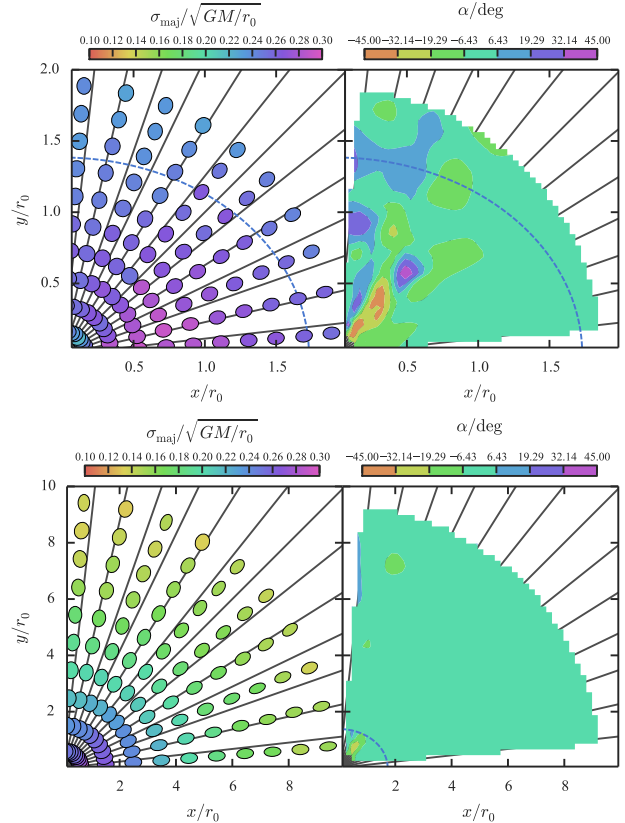

$$\includegraphics[width=\columnwidth]{{{stellarhalo_tracer_beta_nweparams_walterres.nbody_tilt_xy_zoom}}}$$
\vspace{-.84cm}
$$\includegraphics[width=\columnwidth]{{{stellarhalo_tracer_beta_nweparams_walterres.nbody_tilt_xy}}}$$
\caption{Tilt of the stellar halo for $(b/a)_{\mathrm{NFW}}=0.9$, $(c/a)_{\mathrm{NFW}}=0.8$ model in the $(x,y)$ plane: the left
  panels show the velocity ellipses coloured by the magnitude of their
  major axis. The size of the ellipses is unimportant and chosen for
  ease of visualization. The black lines are radial. The right panels
  show the misalignment of the major axis of the ellipses from radial.
  The top two panels correspond to
  $0\lesssim r/r_0 <2$ whilst the lower two correspond to $0\lesssim
  r/r_0 <10$. In all panels the dotted blue line corresponds to
  $q_{\mathrm{S}}=r_{0,\mathrm{S}}$.}
\label{fig::m2m_xy}
\end{figure}

The resulting simulation is plotted in Fig.~\ref{fig::m2m_density}. In
Figures~\ref{fig::m2m_xy},~\ref{fig::m2m_xz} and~\ref{fig::m2m_yz}, we
show the velocity ellipses in the principal planes of the potential
along with the misalignment from spherical alignment. In all three
principal planes, the velocity ellipsoid is misaligned from spherical
by $\lesssim6\degrees$ everywhere outside the scale radius of the
stellar halo. Inside the scale radius, there are also large regions
where the misalignment is $\lesssim6\degrees$ with the largest
misalignments occurring at small $x$ and small $z$ as well as small
$x$ and small $y$. However, inspecting the velocity ellipses in these
regions shows they are very round and so the misalignment measurement
is more susceptible to shot noise. To inspect the impact of shot noise
in the measurement of the tilt from the simulation we show
$\alpha/\sigma_\alpha$ in the $(x,y)$ plane in
Fig.~\ref{fig::m2m_xy_err}. $\sigma_\alpha$ was computed using propagation
of errors from the measured dispersions. We see that the majority of
the plane is consistent with being aligned within the shot noise. This
plane is representative of the other two principal planes and similar
to the simulations shown later. We conclude we are not dominated by
shot noise from the simulation when drawing conclusions on the degree
of alignment. Instead when comparing these models with the data we are
limited by the errors in the observational data.

For the simulation without the anisotropy constraint, we also found
that there were large regions of spherical alignment particularly
outside the scale radius in the $(y,z)$ plane and along the $x$
axis. However, near the $y$ and $z$ axis the ellipses became more
circular with the suggestion that a minor axis is aligned with
spherical polars. Increasing the anisotropy has the effect of
increasing the volume in which the major axis of the velocity
ellipsoid is spherically aligned. We are able to produce a realistic
model of the stellar halo in a triaxial NFW halo that has a large
volume in which that velocity ellipsoid is aligned with spherical
polars within $\sim6\degrees$.

\begin{figure}
$$\includegraphics[width=\columnwidth]{{{stellarhalo_tracer_beta_nweparams_walterres.nbody_tilt_xz_zoom}}}$$
\vspace{-.84cm}
$$\includegraphics[width=\columnwidth]{{{stellarhalo_tracer_beta_nweparams_walterres.nbody_tilt_xz}}}$$
\caption{Tilt of the stellar halo for $(b/a)_{\mathrm{NFW}}=0.9$, $(c/a)_{\mathrm{NFW}}=0.8$ model in the $(x,z)$ plane: see
  Fig.~\protect\ref{fig::m2m_xy} for details.}
\label{fig::m2m_xz}
\end{figure}
\begin{figure}
$$\includegraphics[width=\columnwidth]{{{stellarhalo_tracer_beta_nweparams_walterres.nbody_tilt_yz_zoom}}}$$
\vspace{-.84cm}
$$\includegraphics[width=\columnwidth]{{{stellarhalo_tracer_beta_nweparams_walterres.nbody_tilt_yz}}}$$
\caption{Tilt of the stellar halo for $(b/a)_{\mathrm{NFW}}=0.9$, $(c/a)_{\mathrm{NFW}}=0.8$ model in the $(y,z)$ plane: see
  Fig.~\protect\ref{fig::m2m_xy} for details.}
\label{fig::m2m_yz}
\end{figure}

\begin{figure}
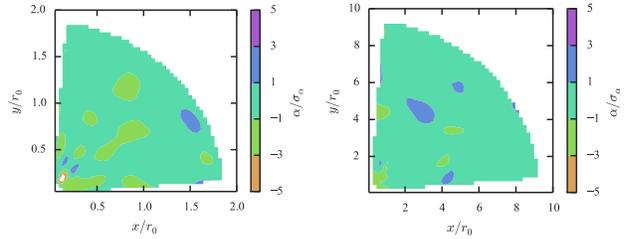

  \centering
  \begin{minipage}{.5\columnwidth}
      \centering
      $$\includegraphics[width=\columnwidth]{{{stellarhalo_tracer_beta_nweparams_walterres.nbody_tilt_xy_zoom_errors}}}$$
  \end{minipage}%
  \begin{minipage}{0.5\columnwidth}
      \centering
      $$\includegraphics[width=\columnwidth]{{{stellarhalo_tracer_beta_nweparams_walterres.nbody_tilt_xy_errors}}}$$
  \end{minipage}
\caption{Tilt of the stellar halo divided by the error in the tilt for $(b/a)_{\mathrm{NFW}}=0.9$, $(c/a)_{\mathrm{NFW}}=0.8$ model in the $(x,y)$ plane. The left panel shows a zoom-in of the right panel. The distributions in the other principal planes are very similar.}
\label{fig::m2m_xy_err}
\end{figure}

The constructed model has a fairly round potential due to the only
weakly triaxial NFW density profile for the dark matter. To
investigate whether the alignment persists for a flatter model we now
go on to construct a model with $(b/a)_{\mathrm{NFW}}=0.7$,
$(c/a)_{\mathrm{NFW}}=0.5$, $(b/a)_{\mathrm{S}}=0.65$ and
$(c/a)_{\mathrm{S}}=0.45$. All other parameters are kept the same. As
expected, the region within which the alignment is less than
$\sim6\degrees$ has decreased compared to the rounder model but there
are still considerable regions, particularly outside $q=r_S$, where
the alignment is spherical. Finally, we report that we constructed a
model with $(b/a)_{\mathrm{NFW}}=1$, $(c/a)_{\mathrm{NFW}}=0.1$,
$(b/a)_{\mathrm{S}}=1$ and $(c/a)_{\mathrm{S}}=0.5$. This model has
spherical alignment at very large radius ($q>6r_S$) and within $q=r_S$
the long axis of the velocity ellipsoid is pointed more towards the
plane for $R>z$ and the short axis is pointed more towards the plane
for $z>R$. However, it should be reported that this model appears to
be only marginally stable as the cost functions drift in time upon
subsequent evolution of the made-to-measure code with no weight
adjustment. Our models do not include the majority of the baryons in
the Galaxy (i.e. the disc) but based on our constructed models we can
make some predictions as to the effects. Including a disc will cause
the velocity ellipses to tilt slightly towards the plane. However, at
large distances the disc potential is dominated by the monopole
component such that this effect will be small for the majority of the
volume studied here. For example, both \cite{Ko10} and
\cite{Bo15} find a flattening for the full potential of
$q\sim0.9$ when fitting the GD-1 stream which lies $\sim15\kpc$ from
the Galactic centre so the models presented here are perhaps already
too flattened \emph{without} including the Galactic disc.

\begin{figure}
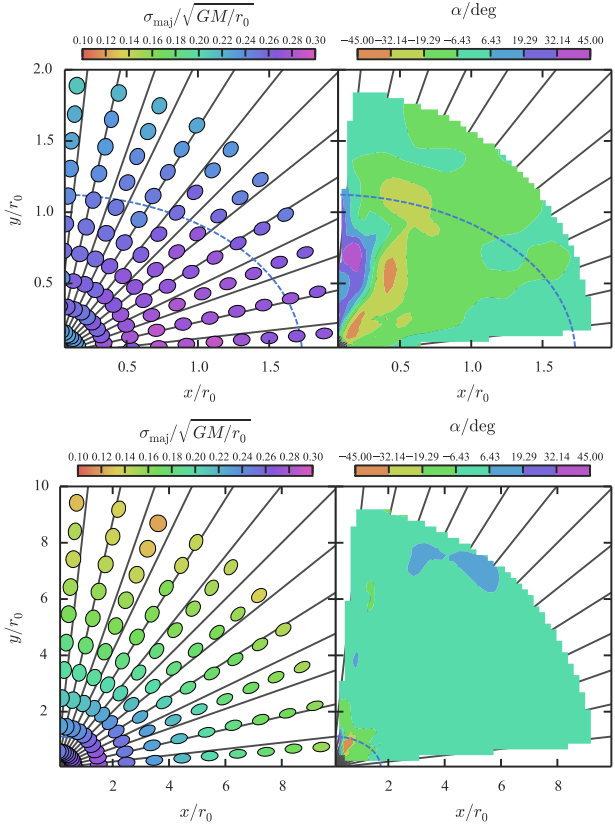

$$\includegraphics[width=\columnwidth]{{{stellarhalo_tracer_beta_nweparams_flatter_walterres.nbody_tilt_xy_zoom}}}$$
\vspace{-.84cm}
$$\includegraphics[width=\columnwidth]{{{stellarhalo_tracer_beta_nweparams_flatter_walterres.nbody_tilt_xy}}}$$
\caption{Tilt of the stellar halo tilt for flatter $(b/a)_{\mathrm{NFW}}=0.7$, $(c/a)_{\mathrm{NFW}}=0.5$ model in the $(x,y)$ plane: see
  Fig.~\protect\ref{fig::m2m_xy} for details.}
\label{fig::m2m_xy2}
\end{figure}
\begin{figure}
$$\includegraphics[width=\columnwidth]{{{stellarhalo_tracer_beta_nweparams_flatter_walterres.nbody_tilt_xz_zoom}}}$$
\vspace{-.84cm}
$$\includegraphics[width=\columnwidth]{{{stellarhalo_tracer_beta_nweparams_flatter_walterres.nbody_tilt_xz}}}$$
\caption{Tilt of the stellar halo for flatter $(b/a)_{\mathrm{NFW}}=0.7$, $(c/a)_{\mathrm{NFW}}=0.5$ model in the $(x,z)$ plane: see
  Fig.~\protect\ref{fig::m2m_xy} for details.}
\label{fig::m2m_xz2}
\end{figure}
\begin{figure}
$$\includegraphics[width=\columnwidth]{{{stellarhalo_tracer_beta_nweparams_flatter_walterres.nbody_tilt_yz_zoom}}}$$
\vspace{-.84cm}
$$\includegraphics[width=\columnwidth]{{{stellarhalo_tracer_beta_nweparams_flatter_walterres.nbody_tilt_yz}}}$$
\caption{Tilt of the stellar halo tilt for flatter $(b/a)_{\mathrm{NFW}}=0.7$, $(c/a)_{\mathrm{NFW}}=0.5$ model in the $(y,z)$ plane: see
  Fig.~\protect\ref{fig::m2m_xy} for details.}
\label{fig::m2m_yz2}
\end{figure}

Notice that we did not introduce any requirement in the cost
function~(\ref{eq:cost}) that drives the made-to-measure solutions to
spherical alignment. We attempted to minimise the tilt angles from
spherical alignment in spherical polar bins beyond a minimum
elliptical radius but this produced no significant change to the
structure of the models. It appears that the alignment cannot be
significantly altered once the potential and tracer density have been
specified. Making the models more radially anisotropic acts to make
the alignment more obvious as the calculation of the tilt is less
susceptible to Poisson noise. Flattened models with strong radial
anisotropy seem to produce near-spherical alignment of the velocity
ellipsoid in significant portions of configuration space without much
difficulty. In other words, the inferred alignment of the velocity
ellipsoid in relatively small spatial volumes does not constrain the
Galactic potential. Strong inferences can only made when global
alignment is detected.  Only with the advent of Gaia proper motions
will we possess datasets of a large enough extent to use the velocity
ellipsoid as a tool for inference on the symmetries of the Galactic
matter distribution.

\section{Conclusions}

This paper has identified the circumstances under which the second
velocity moment tensor of a stellar population is everywhere aligned
in the spherical polar coordinate system.  Exact alignment in
spherical polars is possible if (i) all three semiaxes are the same
and the distribution function (DF) is isotropic, (ii) two of the
semiaxes of the same ($\langle v_r^2 \rangle = \langle v_\theta^2
\rangle$) and the potential is axisymmetric with a DF depending on the
energy $E$ and angular momentum parallel to the symmetry axis $L_z$,
(iii) all three semiaxes are in general different and the potential is
of separable or St\"ackel form in spherical polar coordinates.  In the
latter instance, if the potential is everywhere non-singular, then it
must be spherically symmetric. Our proof is based on the ideas
sketched out in \citet{Sm09a}. An assumption in the proof is that the
even part of the DF is symmetric under time reversal in each
coordinate (i.e., under $v_r \rightarrow -v_r$ or $v_\theta
\rightarrow -v_\theta$ or $v_\phi \rightarrow -v_\phi$).

We have shown that this theorem holds more generally. If the second
velocity moment ellipsoid of a stellar system is triaxial and points
along orthogonal coordinate surfaces, then the potential must be of
separable or St\"ackel form in confocal ellipsoidal coordinates or one
of its limits. The converse of this theorem -- namely, if the
potential is of St\"ackel form then the second velocity moment tensor
aligns in the separable coordinate system -- has been known since the
time of \citet{Ed15}. {\it In our proof, we emphasise that no
  assumption has been made about the form of the DF, or the existence
  of any integrals of the motion quadratic in the velocities.}  All we
have assumed is that the even part of the distribution function
remains invariant under the separate transformations $v_i \rightarrow
-v_i$ in each component (see also \citep{An15}).

It is worth returning to Eddington's (1915) work on the ellipsoidal
hypothesis to spell out the differences. Eddington postulated the
existence of integrals of motion that are quadratic in the velocities.
The ellipsoidal hypothesis is the statement that the DF depends on a
single linear combination of these integrals. Eddington then showed
that the potential has to be of separable or St\"ackel form and the
velocity ellipsoid diagonalises in the separating
coordinates. However, integrals of the motion quadratic in the
velocities only arise from separation of the Hamilton-Jacobi
equation~\citep[see e.g.,][]{Ma67, Ev90}. Hence, Eddington made an
assumption of quadratic integrals that is tantamount to assuming the
potential is of separable form in the first place.

One consequence of our theorem is that the elegant way of solving the
Jeans equation for triaxial velocity ellipsoids using cylindrical
polar alignment developed by \citet{Ca08} and known as ``JAM
modelling'' may only yield physical models if the potential is
separable in the cylindrical polar coordinates. Unfortunately, this
corresponds to total matter distributions that are unlike elliptical
galaxies. If the potential is not of separable form in cylindrical
polars, then there may be galaxy models with DFs that generate
cylindrical alignment with triaxial second velocity moment tensors,
though this however remains to be demonstrated.  JAM models may make
good provisional starting points for constructing N-body or
made-to-measure models, but they are not trustworthy on their own.
Even so, JAM models have had many successes in reproducing global
properties like inclination and mass-to-light ratios in fast rotating
ellipticals~\citep{La12, Dr15}.

What then are the consequences for the stellar halo of the Milky Way?
Our theorem applies to stellar systems in which the velocity ellipsoid
is spherically or spheroidally aligned everywhere. The data on halo
stars of the Milky Way do suggest it is aligned with the spherical
polar coordinate system (usually within $3^\circ$) for Galactocentric
radii between $\sim 6$ and $\sim 11$ kpc in the Northern
hemisphere. There are one or two bins with more substantial
deviations, though this may be partly a consequence of contamination
by the thick disk at low latitude. We have also shown that alignment
in prolate spheroidal coordinates does not give a markedly better fit
than spherical alignment. In both calculations, no attempt has been
made to remove substructure (such as Sagittarius stream stars or tidal
debris) from the stellar halo sample. Such substructure is of course
expected in $\Lambda$CDM and might spoil any exact alignment for an
underlying smooth halo population.

\cite{Bi11} have argued that only limited inferences on the potential
may be drawn from the orientation of the second velocity moment
ellipsoid. They constructed DFs of the thin and thick disks of the
Galaxy using orbital tori. They showed that at some locations above
the plane in the vicinity of the Sun, the velocity ellipsoid is
spherically aligned despite the matter distribution being highly
flattened.  However, the more recent data on the stellar halo provided
by \cite{Bo10} does provide a more substantial challenge. Now the
alignment is known to be almost spherical over a large swathe of the
Galaxy, and it is unclear whether flattened models exist that can
provide this.

To understand whether such approximate alignment is consistent with
flattened or triaxial potentials, we have used made-to-measure
modelling~\citep{Sy96, Dehnen2009}. We constructed a triaxial stellar
halo tracer model in an triaxial NFW dark matter potential. The tracer
population was chosen to be more triaxial ($b/a=0.8,c/a=0.6$) than the
NFW profile ($b/a=0.9,c/a=0.8$). The stellar halo was chosen to have a
double power-law density profile and an anisotropy profile that went
from isotropic at the centre to very radial in the outskirts. Despite
the triaxiality of the potential, it was found that the velocity
ellipsoid had a major axis that deviated from spherical alignment by
$\lesssim6\degrees$ for large regions of space. Outside the scale
radius of the stellar halo, the velocity ellipsoid is nearly
everywhere spherically aligned. We went on to investigate a model with
a flatter NFW profile ($b/a=0.7,c/a=0.5$) and found that the volume of
the spherical alignment region was decreased within the scale radius
of the model although there were still considerable regions where the
alignment was spherical.

If the alignment has to be exactly spherical everywhere, then the
restriction on the potential is very severe. The potential has to be
spherical if it is everywhere non-singular. However, if the alignment
is only close to spherical over a substantial portion of configuration
space, then a much greater variety of potentials are possible,
including flattened ones. Strong inference on the potential can only
be made if global spherical alignment is established. The Gaia
satellite will provide six-dimensional phase space information for
stars brighter than $V\approx 20$ within 20 kpc of the Sun. This may
provide datasets of sufficient global coverage to enable the alignment
of the velocity ellipsoid to be used as a tool for constraining the
Galactic potential.

\section*{Acknowledgments}
NWE thanks Zeljko Ivezic for stimulating him to think about this
matter again, as well as providing data files and advice on the
propagation of errors. JLS acknowledges the use of the python package
\textsc{pynbody} \citep{Pontzen} for the visualization of $N$-body
snapshots. We thank an anonymous referee for a helpful report.

\bibliographystyle{mn2e}

\appendix

\section{Eddington Potentials without Tears}

The separable potentials in spherical polars
\begin{equation}
\psi = \psi_1(r) + {\psi_2(\theta)\over r^2} + {\psi_3(\phi)\over
  r^2\sin^2\theta},
\label{eq:eddpot}
\end{equation}
were introduced into stellar dynamics by Eddington. They are often
called Eddington potentials in the astronomical
literature~\citep[e.g.,][]{Cl37,LB62,Lu87}. Astrophysically useful
potentials must have $\psi_3(\phi) =0$ to avoid singularities all
along the polar axis.

The density then must have form
\begin{equation}
\rho = \rho_0(r) + {g(\theta)\over r^4}.
\end{equation}
This may also seem objectionable as the density has a serious
divergence at the origin.  In fact, Eddington (1915) himself believed
that $\psi_2(\theta) = 0$ and so $g(\theta)=0$ for practical
solutions.  However, this is not the case, as the awkwardness can be
avoided by requiring the non-spherical part of the density to fall
like $r^{-4}$ only outside the central region.

We now give a simple example of an axisymmetric density distribution
that has such a property. We choose the density as
\begin{equation}
\rho(r,\theta) = \rho_0(r) + \rho_2(r) P_2(\cos\theta),
\end{equation}
where $P_2$ is the Legendre polynomial. For flattened systems with
everywhere positive mass density, we need $\rho_2 \leq 0$ and $\rho_0(r) + \rho_2(r)\ge 0$
for all $r > 0$. The gravitational potential of such systems is given by
\begin{equation}
\psi = \psi_0(r) + h(r) P_2(\cos\theta).
\end{equation}
where
\begin{equation}
{\psi_0(r) \over 4\pi G} = {1\over r} \int_0^r r^2\rho_0(r) dr +
\int_r^\infty\rho_0(r)dr
\end{equation}
and
\begin{equation}
{5\over 4\pi G}h(r) = {1\over r^3}\int_0^r r^4 \rho_2(r) dr +
r^2\int_r^\infty \rho_2(r) {dr\over r},
\end{equation}
We wish to have $h \propto r^{-2}$ for $r\ge r_0$. In this
region, we need to set $\rho_2(r) = \rho_2(r_0) (r_0/r)^4$ to ensure a
continuous density.

We now define $K$ to be
\begin{equation}
K = \int_0^{r_0} \left[r^4\rho_2(r) - r_0^4 \rho_2(r_0)\right]dr
\end{equation}
and we evaluate our formula for $h$ at $r>r_0$ to get
\begin{equation}
{h(r) \over 4 \pi G} = {K\over 5r^3} + {r_0^4 \rho_2(r_0)\over 4 r^2}
\end{equation}
Evidently, the potential is of the desired form for $r>r_0$ provided
that $\rho_2(r)$ for $r< r_0$ obeys the simple integral constraint
that $K=0$. Only for $r>r_0$ is the potential of separable form. Orbits
whose pericenters satisfy $r_{\rm p} > r_0$ lie within rectangular
toroids and have three exact integrals of motion.

\section{Alignment in Ellipsoidal Coordinates}

Ellipsoidal coordinates ($\lambda,\mu, \nu$) are the most natural
coordinates to study triaxial stellar
systems~\citep[e.g.,][]{LB62,DZ85,vdV,Sa15}. The scale factors are
\begin{eqnarray}
P^2 &=& {(\lambda-\mu)(\lambda-\nu) \over 4(\lambda + a)(\lambda+b)(\lambda+c)},\nonumber\\
Q^2 &=& {(\mu -\lambda)(\mu-\nu) \over 4(\mu + a)(\mu+b)(\mu+c)},\nonumber \\
R^2 &=& {(\nu -\mu)(\nu-\lambda) \over 4(\nu + a)(\nu+b)(\nu+c)},
\label{eq:scalefacsellips}
\end{eqnarray}
where $a$, $b$ and $c$ are constants that define two
sets of foci. Surfaces of constant $\lambda$, $\mu$ and $\nu$ are
confocal ellipsoids, one-sheeted hyperboloids and two-sheeted
hyperboloids respectively. The coordinate system is illustrated in
e.g., \citet{Hi99}, \citet{DZ85} or \citet{Bo96}.

We start by assuming that the second velocity moment ellipsoid is
diagonalised in confocal ellipsoidal coordinates, so that all the
cross-terms $\langle v_\lambda v_\mu \rangle$, $\langle v_\lambda
v_\nu \rangle$ and $\langle v_\mu v_\nu \rangle$ all vanish.  The
Hamiltonian is:
\begin{equation}
H = {1\over 2} \Bigl( {p_\lambda^2\over P^2} + {p_\mu^2 \over Q^2} +
{p_\nu^2 \over R^2} \Bigr) - \psi(\lambda,\mu,\nu),
\end{equation}
where $p_\lambda = P^2 {\dot \lambda}$, $p_\mu = P^2 {\dot \mu}$ and
$p_\nu = P^2 {\dot \nu}$ are canonical momenta. Again, as the
Hamiltonian is time-invariant, we can always construct the even part
of the DF and assume that it depends on the squares of the canonical
momenta only. For the cross-terms to vanish in ellipsoidal
coordinates, this means $\Fe = \Fe(p_\lambda^2, p_\mu^2,
p_\nu^2,\lambda,\mu,\nu)$. Using the Hamiltonian to eliminate
$p_\lambda^2$ without loss of generality, we obtain $\Fe(H,p_\mu^2,
p_\nu^2,\lambda,\mu,\nu)$. Taking the Poisson bracket $\{\Fe,H\}=0$
tells us that $\Fe$ is independent of $\lambda$ as well as
$p_\lambda$. We find, after some work,:
\begin{equation}
A_1 {\partial \Fe \over \partial p_\mu} + A_2 {\partial \Fe \over
  \partial p_\nu} = {p_\mu\over Q^2} {\partial \Fe \over \partial
  \mu} + {p_\nu\over R^2} {\partial \Fe \over \partial \nu}
\end{equation}
with
\begin{equation}
A_1 = {1\over \lambda-\mu}\left[{p_\mu^2\over 2}{\partial \over
    \partial \mu} \left( {\lambda - \mu \over Q^2} \right)
+
{p_\nu^2\over 2}{\partial \over
    \partial \mu} \left( {\lambda - \mu \over R^2} \right)
-{\partial \over \partial \mu} \left( (\lambda-\mu) \psi \right) + H
\right]
\end{equation}
and
\begin{equation}
A_2 = {1\over \lambda-\nu}\left[{p_\mu^2\over 2}{\partial \over
    \partial \nu} \left( {\lambda - \nu \over Q^2} \right)
+
{p_\nu^2\over 2}{\partial \over
    \partial \nu} \left( {\lambda - \nu \over R^2} \right)
-{\partial \over \partial \nu} \left( (\lambda-\nu) \psi \right) + H
\right]
\end{equation}
We are at liberty to send $p_\mu \rightarrow -p_\mu$ or to send $p_\nu
\rightarrow -p_\nu$ as the DF is invariant under such changes. This
tell us that the two equations
\begin{eqnarray}
A_1  {\partial \Fe \over \partial p_\mu} &=&
{p_\mu\over Q^2} {\partial \Fe \over \partial \mu},\nonumber\\
A_2  {\partial \Fe \over \partial p_\nu} &=&
{p_\nu\over R^2} {\partial \Fe \over \partial \nu},
\end{eqnarray}
must be separately satisfied. Now, take the first equation, multiply
by $(\lambda-\mu)$ and differentiate with respect to $\lambda$. We
already know that $\Fe$ is independent of $\lambda$ and that $H$ may
be treated as a constant, so this operator annihilates all terms bar
the one containing the potential and leaves us with
\begin{equation}
{\partial^2\over \partial\lambda\partial\mu} \Bigl( (\lambda-\mu)\psi
\Bigr) = 0.
\end{equation}
Similarly, multiplying the second equation by $(\lambda-\nu)$ and
differentiating with respect to $\lambda$ leaves us with
\begin{equation}
{\partial^2\over \partial\lambda\partial\nu} \Bigl( (\lambda-\nu)\psi
\Bigr) = 0.
\end{equation}
We could of course have started by eliminating $p_\mu^2$ in terms of
the Hamiltonian, and repeating our steps would yield
\begin{equation}
{\partial^2\over \partial\mu\partial\nu} \Bigl( (\mu-\nu)\psi
\Bigr) = 0.
\end{equation}
This gives us three partial differential equations that the potential
must satisfy, and it is straightforward to integrate them up to
establish
\begin{equation}
\psi(\lambda,\mu,\nu) =
{f_1(\lambda) \over (\lambda  -\mu)(\lambda-\nu)} +
{f_2(\mu) \over (\mu -\lambda)(\mu-\nu)} +
{f_3(\nu) \over (\nu - \lambda)(\nu -\mu)}
\end{equation}
where $f_1(\lambda)$, $f_2(\mu)$ and $f_3(\nu)$ are arbitrary
functions of the indicated arguments. This is the separable or
St\"ackel potential in confocal ellipsoidal coordinates.

\section{The St\"ackel Condition}

Rather than demonstrating the theorem for each alignment separately, a
more mathematical -- but abstract -- approach is to derive all
possible coordinate systems and gravitational potentials
together. This is similar in spirit to the original investigations of
\citet{St91} and \citet{Ed15}.

Here, we prove the following theorem (see also An \& Evans 2015): {\it
  Suppose that (i) the second velocity moment tensor of a stellar system
  is aligned in an orthogonal curvilinear coordinate system and has
  (in general) three unequal axes and that (ii) the stellar system is
  in a steady state, so that the even part of the distribution
  function (DF) satisfies the collisionless Boltzmann equation and
  (iii) the DF is invariant under reversal of the sign of each
  velocity component. Then, it necessarily follows that the coordinate
  system is the confocal ellipsoidal coordinates (or one of its
  limiting cases) and that the gravitational potential is of separable
  or St\"ackel form.}

Consider a system with 3 degrees of freedom governed by the
Hamiltonian of the form of
\begin{equation}
H=\frac12\sum_k\frac{p_k^2}{\mathsf h_k^2(q_1,q_2,q_3)}
-\psi(q_1,q_2,q_3),
\end{equation}
where ($q_1,q_2,q_3$) are orthogonal curvilinear coordinates,
($p_1,p_2,p_3$) are the corresponding canonical momenta and $h_1, h_2,
h_3$ are the scale factors. Suppose that the system admits an integral
of motion of the form, $\Fe=\Fe(p_1^2,p_2^2,p_3^2;q_1,q_2,q_3)$ which
is recognised as the even part of the DF. The vanishing of the Poisson
bracket requires
\begin{eqnarray}
\dot{\Fe}
&=&\left\lbrace \Fe,H\right\rbrace
=\sum_i\left(\frac{\partial\Fe}{\partial q_i}
\frac{\partial H}{\partial p_i}
-\frac{\partial H}{\partial q_i}
\frac{\partial\Fe}{\partial p_i}\right)\nonumber \\
&=&\sum_ip_i\left(\frac1{\mathsf h_i^2}\frac{\partial\Fe}{\partial q_i}
-2\frac{\partial H}{\partial q_i}
\frac{\partial\Fe}{\partial(p_i^2)}\right)=0.
\end{eqnarray}
However, both $H$ and $\Fe$ are invariant under $p_j\to-p_j$ for any
$j$'s, and so it follows that, for ${\forall}i\in\{1,2,3\}$,
\begin{equation}
\frac{\partial\Fe}{\partial q_i}
=\zeta_i\frac{\partial H}{\partial q_i}
,\quad
\zeta_i\equiv
2\mathsf h_i^2\frac{\partial\Fe}{\partial(p_i^2)}.
\end{equation}
Here note that, for any $i$ and $j$,
\begin{eqnarray}
\frac{\partial\zeta_i}{\partial q_j}
&=& 2\mathsf h_i^2\frac{\partial^2\Fe}{\partial q_j\partial(p_i^2)}
+2\frac{\partial\mathsf h_i^2}{\partial q_j}
\frac{\partial\Fe}{\partial(p_i^2)}\nonumber\\
&=&2\mathsf h_i^2\frac{\partial}{\partial(p_i^2)}
\left(\zeta_j\frac{\partial H}{\partial q_j}\right)
+\frac{\zeta_i}{\mathsf h_i^2}\frac{\partial\mathsf h_i^2}{\partial
  q_j}\nonumber \\
&=& 2\mathsf h_i^2\frac{\partial\zeta_j}{\partial(p_i^2)}
\frac{\partial H}{\partial q_j}
+2\mathsf h_i^2\zeta_j\frac{\partial^2 H}{\partial(p_i^2)\partial q_j}
+\frac{\zeta_i}{\mathsf h_i^2}\frac{\partial\mathsf h_i^2}{\partial
  q_j}\nonumber \\
&=&4\mathsf h_i^2\mathsf h_j^2
\frac{\partial^2 \Fe}{\partial(p_i^2)\partial(p_j^2)}
\frac{\partial H}{\partial q_j}
+\mathsf h_i^2\zeta_j
\frac\partial{\partial q_j}\left(\frac1{\mathsf h_i^2}\right)
+\frac{\zeta_i}{\mathsf h_i^2}\frac{\partial\mathsf h_i^2}{\partial q_j}\nonumber\\
&=& 4\mathsf h_i^2\mathsf h_j^2
\frac{\partial^2 \Fe}{\partial(p_i^2)\partial(p_j^2)}
\frac{\partial H}{\partial q_j}
+\frac{\zeta_i-\zeta_j}{\mathsf h_i^2}\frac{\partial\mathsf h_i^2}{\partial q_j}
\end{eqnarray}
Then the integrability condition on $\Fe$ is
\begin{equation}
\frac\partial{\partial q_i}\left(\frac{\partial\Fe}{\partial q_j}\right)
-\frac\partial{\partial q_j}\left(\frac{\partial\Fe}{\partial q_i}\right)
=0
\end{equation}
which results in
\begin{eqnarray}
&& \frac\partial{\partial q_i}
\left(\zeta_j\frac{\partial H}{\partial q_j}\right)
-\frac\partial{\partial q_j}
\left(\zeta_i\frac{\partial H}{\partial q_i}\right)\nonumber\\
&=&\frac{\partial\zeta_j}{\partial q_i}\frac{\partial H}{\partial q_j}
-\frac{\partial\zeta_i}{\partial q_j}\frac{\partial H}{\partial q_i}
+\zeta_j\frac{\partial^2 H}{\partial q_i\partial q_j}
-\zeta_i\frac{\partial^2 H}{\partial q_j\partial q_i}\nonumber\\
&=&\frac{\zeta_j-\zeta_i}{\mathsf h_j^2}
\frac{\partial\mathsf h_j^2}{\partial q_i}
\frac{\partial H}{\partial q_j}
-\frac{\zeta_i-\zeta_j}{\mathsf h_i^2}
\frac{\partial\mathsf h_i^2}{\partial q_j}
\frac{\partial H}{\partial q_i}
+(\zeta_j-\zeta_i)\frac{\partial^2 H}{\partial q_i\partial q_j}\nonumber\\
&=&(\zeta_j-\zeta_i)\,\mathcal D_{ij}(H)=0
\quad\text{(for all $i,j$)}.
\end{eqnarray}
Here, $\mathcal D_{ij}(f)$ is the linear second-order differential operator
acting on a function $f(q_1,q_2,q_2)$, defined as
\begin{eqnarray}
\mathcal D_{ij}(f) &\equiv&
\frac1{\mathsf h_j^2}\frac{\partial\mathsf h_j^2}{\partial q_i}
\frac{\partial f}{\partial q_j}
+\frac1{\mathsf h_i^2}\frac{\partial\mathsf h_i^2}{\partial q_j}
\frac{\partial f}{\partial q_i}
+\frac{\partial^2f}{\partial q_i\partial q_j}\nonumber\\
&=&\left(\frac{\partial\ln\mathsf h_j^2}{\partial q_i}\frac\partial{\partial q_j}
+\frac{\partial\ln\mathsf h_i^2}{\partial q_j}\frac\partial{\partial q_i}
+\frac{\partial^2}{\partial q_i\partial q_j}\right)f,
\end{eqnarray}
which is symmetric for $i\leftrightarrow j$, i.e.\ $\mathcal
D_{ij}(f)=\mathcal D_{ji}(f)$.  In other words, if there exists an
integral $\Fe$, we must have $\zeta_i=\zeta_j$ or $\mathcal
D_{ij}(H)=0$ for any pair of indices $i$ and $j$. The
$\zeta_i=\zeta_j$ (for $i\ne j$) case however implies that the
integral $\Fe$ becomes invariant under the rotation within
$p_i$-$p_j$ plane and so the distribution must be isotropic within
$q_i$-$q_j$ plane: that is to say, the resulting second velocity moments
must be degenerate as in $\langle v_i^2\rangle=\langle
v_j^2\rangle$. If $\zeta_i\ne\zeta_j$ on the other hand, we must have
\begin{equation}
\mathcal D_{ij}(H)
=\frac12\sum_k\mathcal D_{ij}\!\left(\frac1{\mathsf h_k^2}\right)\,p_k^2
-\mathcal D_{ij}(\psi)=0
\quad\Rightarrow\
\mathcal D_{ij}(\mathsf h_k^{-2})=\mathcal D_{ij}(\psi)=0
\end{equation}
The condition on the scale factors $\mathcal D_{ij}(\mathsf
h_k^{-2})=0$ for all $i\ne j$ (and any $k$) is the same condition
defining the St\"ackel systems. The most general orthogonal
curvilinear coordinate in a Euclidean space that satisfies the
condition is the confocal ellipsoidal coordinates. This encompasses
the 11 3-D quadric coordinates in which the Helmholtz equation
separates~\citep{MF}.  On the other hand, the general solution of
$\mathcal D_{ij}(\psi)=0$ in the confocal ellipsoidal coordinates (or
its degenerate limit) is known to be
$\psi(q_1,q_2,q_3)=\sum_kf_k(q_k)/\mathsf h_k^2$ where $f_k(q_k)$
is an arbitrary function of the coordinate component $q_k$ alone.

The condition $\mathcal D_{ij}(\psi)=0$ is really the
integrability condition on the system of the quasi-linear partial
differential equations. If we suppose the existence of the set of
functions $\{f_k(q_k)\}$ such that
$\psi(q_1,q_2,q_3)=\sum_kf_k(q_k)/\mathsf h_k^2$, then
\[
\frac{\partial\psi}{\partial q_i}
=\frac{f_i'(q_i)}{\mathsf h_i^2}
-\sum_k\frac{\partial\mathsf h_k^{-2}}{\partial q_i}f_k(q_k)
\quad\Rightarrow\
\frac{\partial f_i}{\partial q_j}
=\delta_i^j\mathsf h_i^2\left(\frac{\partial\psi}{\partial q_i}
+\sum_k\frac{\partial\mathsf h_k^{-2}}{\partial q_i}f_k\right).
\]
where $\delta_i^j$ is the Kronecker delta.  This is the system of
partial differential equations on $\{f_k(q_1,q_2,q_3)\}$, whose
compatibility condition implies that $(\partial/\partial q_j)(\partial
f_i/\partial q_k) =(\partial/\partial q_k)(\partial f_i/\partial q_j)$
for any $i,j,k$. The only non-trivial conditions among these are
\[
\frac{\partial}{\partial q_j}\left(\frac{\partial f_i}{\partial q_i}\right)
=\mathsf h_i^2\left[\mathcal D_{ij}(\psi)
-\sum_k\mathcal D_{ij}(\mathsf h_k^{-2})f_k\right]=0
\quad\text{(for any $i\ne j$)},
\]
and so, given the St\"ackel coordinate satisfying $\mathcal
D_{ij}(\mathsf h_k^{-2})=0$, we find that $\mathcal D_{ij}(\psi)=0$ is
the necessary condition for existence of the solution set
$\{f_k\}$. Moreover, thanks to the Frobenius theorem, the condition is
also sufficient for (local) existence of such a solution set. In other
words, $\psi(q_1,\dotsc,q_n)=\sum_kf_k(q_k)/\mathsf h_k^2$ is also in
fact the general solution of $\mathcal D_{ij}(\psi)=0$.

A consequence of this theorem is that it seems to imply that the only
axisymmetric equilibria with DFs are either (i) Jeans models with $F =
F(E,L_z)$ in which two of the semi-axes of the second moment tensor
are the same or (ii) St\"ackel or separable models in spheroidal
coordiantes with all three semiaxes different. This seems to cast
doubt on the existence of axisymetric equilibria constructed by
Schwarzschild modelling~\citep[see e.g.,][]{Cr99}. However, the likely
resolution of this paradox is that the only regular orbits observing
the required symmetry are those in separable potentials, and so the
velocity distributions resulting from a Schwarzschild model in a
non-separable potential are not strictly symmetric.

\end{document}